\documentclass[iop,apj]{emulateapj}
\usepackage{amsmath,amssymb,amstext}

\usepackage[breaklinks,colorlinks,citecolor=blue,linkcolor=magenta]{hyperref} 

\usepackage[all]{hypcap} 

\usepackage{amssymb}
\usepackage{amsmath}
\usepackage{graphicx}

\newcommand{\bnabla}{{\mbox{\boldmath$\nabla$}}}

\usepackage{aas_macros}
\usepackage{natbib}
\shorttitle{Cluster heating by cosmic rays}
\shortauthors{Ruszkowski, Yang, and Reynolds}

\begin{document}

\title{Cosmic ray feedback heating of the intracluster medium}
\author{Mateusz Ruszkowski$^{1,2}$, H.-Y. Karen Yang$^{2,3}$, and Christopher S. Reynolds$^{2}$}
\affil{$^{1}$Department of Astronomy, University of Michigan, 1085 S University Ave, 311 West Hall, Ann Arbor, MI 48109}
\affil{$^{2}$Department of Astronomy, University of Maryland, College Park, MD 20742}
\altaffiltext{3}{{\it Einstein} Fellow}
\email{mateuszr@umich.edu (MR), hsyang@astro.umd.edu (KY), chris@astro.umd.edu (CR)}

\begin{abstract}
Self-regulating active galactic nuclei (AGN) feedback in the cool cores of galaxy clusters plays central role in solving the decades-old cooling flow problem. 
While there is consensus that AGN provide most if not all of the energy needed to offset radiative losses in the intracluster medium (ICM) and prevent catastrophically large star formation rates, one major problem remains unsolved -- how is the AGN energy thermalized in the ICM and what are the effective black hole feeding rates in realistic systems? We perform a suite of three-dimensional magneto-hydrodynamical (MHD) adaptive mesh refinement simulations of AGN feedback in a cool core cluster including cosmic ray (CR) physics. CRs are supplied to the ICM via collimated AGN jets and subsequently disperse in the magnetized ICM via streaming, and interact with the ICM via hadronic, Coulomb, and streaming instability heating. We find that CR transport is an essential model ingredient needed for AGN feedback to self-regulate, at least within the context of the physical model considered here. When CR streaming is neglected, the suppression of CR mixing with the ICM by magnetic fields significantly reduces ICM heating, which leads to cooling catastrophes.  In the opposite case, when CR streaming is included, CRs come into contact with the ambient ICM and efficiently heat it, which results in globally stable atmospheres. Moreover, the dynamical state and intermittency of the central AGN are dramatically altered when CR streaming is present -- while the AGN is never in a completely off-state, it is more variable, and the atmosphere goes through cycles characterized by low gas velocity dispersion interspersed with more violent episodes. We find that CR streaming heating dominates over the heating due to Coulomb and hadronic processes. Importantly, in simulations that include CR streaming, CR pressure support in the central 100 kpc is very low and does not demonstrably violate observational constraints. On the contrary, when CR streaming is neglected, CR energy is not spent on the ICM heating and CR pressure builds up to the level that is in disagreement with the data. Overall, our models demonstrate that CR heating is a viable channel for the thermalization of AGN energy in clusters, and likely also in elliptical galaxies, and that CRs play an important role in determining AGN intermittency and the dynamical state of cool core atmospheres.
\end{abstract}
 
\keywords{cosmic rays -- galaxies: active -- galaxies: clusters: intracluster medium}
\maketitle

\section{Introduction}
One of the long-standing puzzles in modeling of galaxy clusters is the ``cooling-flow problem'' \citep{Fabian1994} -- clusters with short central radiative cooling times, i.e., cool-core clusters, are predicted to host massive inflows of gas and to harbor large amounts of cold gas and stars near their centers, significantly in excess of what is observed. Various heating mechanisms of the ICM in cool cores have been proposed in order to prevent or reduce these inflows, among which AGN feedback is the most promising one \citep{McNamaraNulsen2012}. These mechanisms include heating by dynamical friction acting on substructure (e.g., \citet{elzant2004}), 
conduction of heat from the outer hot  layers of cool cores to their centers ({e.g., \citet{BalbusReynolds2008}, \citet{Bogdanovic2009}, \citet{Parrish2010, RuszkowskiOh2010, ZakamskaNarayan2003, RuszkowskiOh2011,Ruszkowski2011}),  precipitation-driven AGN feedback (e.g., \citet{GaspariRuszkowskiSharma2012, Li2015, Li2016}), conduction and AGN feedback (e.g., \citet{RuszkowskiBegelman2002, YangReynolds2016a}),  dissipation of AGN-induced sound waves and weak shocks (e.g., \citet{Fabian2003, Li2016, Ruszkowski2004a,Ruszkowski2004b,Fabian2017}), and cosmic ray heating (e.g., \citet{GuoOh2008}). Strong argument in favor of the AGN mechanism comes from the prevalence of AGN jet-inflated radio bubbles in cool-core clusters and the correlation between the estimated jet power and central cooling luminosity. Despite the observational evidence supporting AGN feedback, numerical modeling of AGN accretion and feedback suffers from large uncertainties rooted in the huge separation of scales between the size of supermassive black hole accretion disks and that of clusters. Another major unsolved problem in modeling AGN feedback concerns the issue of thermalization of the AGN jet energy in the ICM. Detailed understanding of this process is needed to discover how the supermassive black hole feedback and feeding really work in realistic systems. \\
\indent
In recent years hydrodynamic simulations made substantial progress in terms of understanding AGN accretion and feedback processes in clusters. Earlier simulations that include Bondi accretion of hot gas and injection of thermal energy demonstrated that supermassive black hole feedback can be self-regulated (e.g., \citet{Sijacki2007}). More recently, motivated by multiple theoretical and observational studies that focus on the role of thermal instability in the ICM in feeding the central supermassive black hole (e.g., \citet{McCourt2012,voit2015}), simulations including cold-gas accretion and momentum-driven feedback have successfully reproduced the positive temperature gradients and properties of cold gas within the cool cores \citep{GaspariRuszkowskiSharma2012, Li2015, Li2016}. These kinds of simulations provided valuable insights into the mysteries of how the AGN energy is transformed into heat and how the heat is distributed radially and isotropically throughout the cool core. Specifically, \citet{YangReynolds2016b} and \citet{Li2016} showed that mixing with ultra-hot thermal gas within bubbles and shock heating are the dominant heating mechanisms. Moreover,  \citet{YangReynolds2016b} showed that a gentle circulation flow on billion-year timescale is responsible for partially compensating cooling and transporting the heat provided by the AGN in an isotropic manner.\\
\indent
Despite these successes, fundamental and important physical processes are not captured in purely hydrodynamic models. One of the assumptions of the above-mentioned hydrodynamic models is that, because the injected kinetic energy is quickly turned into thermal energy by shocks during the initial inflation phase, the bubbles are filled with ultra-hot thermal gas. In reality, the composition of radio bubbles is still largely unknown. Observational estimates generally show that the pressure contributed by radio-emitting CR electrons plus magnetic pressure is small compared to the ambient pressure, suggesting that the bubbles are dominated by either non-radiating CR particles or ultra-hot thermal gas \citep{DunnFabian2004}. While momentum-driven jet models often produce radially elongated bubbles, CR-dominated light jets can naturally inflate fat bubbles like those observed at the center of Perseus \citep{GuoMathews2011}. Both types of bubble shapes appear to exist in observed cool cores, suggesting that the bubbles could have a range of different compositions \citep{Guo2016}. In terms of heating the ICM, CR-dominated bubbles are expected to behave qualitatively differently from hydrodynamic bubbles. First, they expand with an effective adiabatic index of 4/3 instead of 5/3. Second, while mixing is a primary heating mechanism for hydrodynamic bubbles, CR bubbles contain less thermal energy that could be accessed by the ICM via mixing. Also, the level of mixing and the distance bubbles could travel before getting disrupted by instabilities depend on a number of factors, such as the smaller amount of momentum they carry, their lower  density, CR diffusion along the magnetic field, and the topology of the magnetic field in the ICM \citep{Ruszkowski2007, Ruszkowski2008}. Third, the surrounding ICM partially mixed with the CR bubbles is more buoyant and could result in a significant outward mass transfer. In fact, \citet{MathewsBrighenti2008} showed that this has a net cooling effect on the gas as the ICM displaced by the CR bubbles expands. Therefore, it is unclear how the heating occurs and how self-regulation can be established in cases where CRs dominate the bubble energy content. Some recent works on CR bubbles focused on 2D simulations; however, 3D simulations are required in order to accurately capture the properties of mixing.\\
\indent
CRs can scatter on either magnetic field irregularities generated by externally driven turbulence or by self-excited Alfv{\'e}n waves via the CR streaming instability.
In the latter case CRs stream down their pressure gradients along magnetic field lines at (or above) the Alfv{\'e}n speed. In this case, CRs experience an effective drag force that heats the gas \citep{Zweibel2013}. This Alfv{\'e}n wave heating was proposed as a viable mechanism to offset radiative cooling \citep{Loewenstein1991, GuoOh2008, Pfrommer2013, JacobPfrommer2016a, JacobPfrommer2016b}. However, so far only spherically symmetric 1D models of Alfv{\'e}n wave heating were explored in the literature.\\
\indent
In this paper we study the ICM heating by CR-dominated bubbles using 3D MHD simulations including CR advection, streaming, Alfv{\'e}n wave heating due to streaming and CR heating due to hadronic interactions between CRs and the thermal ICM. We demonstrate that CR transport by streaming is essential for constructing self-regulating feedback loop models, at least within the context of the physical model considered here. We show that CR contribution to the heating budget can be very important and that heating due to streaming can dominate over the hadronic and Coulomb heating. We also show that the simulations that include CR heating result in more intermittent AGN feedback. \\
\indent
The paper is organized as follows. In Section 2 we describe basic physics relevant to CR heating of the ICM and the numerical techniques employed in our work. In Section 3 we present our main results. Summary and Conclusions are presented in Section 4. 

\section{Methods}
\subsection{Initial and boundary conditions and the jet feedback model}
The gravitational potential and initial conditions for the temperature and density distributions of the gas resemble those adopted by \citet{YangReynolds2016b}.
In brief, the cluster atmosphere is initially close to hydrostatic equilibrium and its density profile is similar to that corresponding to the Perseus cluster. \\
\indent
We include tangled magnetic fields that are generated using the method similar to that described in \citet{Ruszkowski2007}. We assume that in Fourier space the field has the following form
\begin{equation}
B\propto k^{-11/6}\exp\left[-\left(\frac{k}{k_{\rm in}}\right)^{4} \right]   
\end{equation}
where $k_{\rm in}=10^{2}(2\pi/L)$, 
where $L=1$Mpc is the size of the computational domain. We perform an inverse Fourier transform to generate real-space magnetic fields and, following \citet{wiener2013}, we rescale the field such that  $B\propto \rho_{o}^{0.3}$, where $\rho_{o}$ is the ICM density. This ensures that the magnetic pressure is approximately proportional to the gas pressure. In order to generate divergence-free field, we Fourier transform the field and perform divergence cleaning as in \citet{Ruszkowski2007}. This procedure is repeated until  a divergence-free field proportional to $\rho_{o}^{0.3}$ is obtained. The final field is normalized such that plasma $\beta\sim10^{2}$.  We also impose small isobaric perturbations $\delta\rho/\rho$ on top of the average gas density profiles. Following \citet{Gaspari2012}, these fluctuations are approximately characterized by white noise spectrum with the amplitude of 0.1. The resulting ICM gas density distribution is given by $\rho=\rho_{o}\max(0.8,1+\delta\rho/\rho)$.\\
\indent
We use adaptive mesh refinement to refine the domain up to the maximum resolution of 1.95 kpc. Refinement is triggered by temperature gradients. We employ diode boundary conditions (the gas is only allowed to flow out of the domain; code variables have vanishing gradient at the boundary) but note that the choice of boundary conditions is not critical as the domain is much larger than the size of the central parts of the cool core. \\
\indent
The black hole feedback model adopted here is based on the ``chaotic cold accretion'' model \citep{Gaspari2012,Gaspari2013,Li2015} and closely follows that used by \citet{YangReynolds2016b}. In this model the cooling gas is removed from the hot phase of the ICM when its temperature drops below $T=5\times10^{5}$K. The cold gas is then converted to passive particles that follow the fluid and are allowed to accrete onto the central black hole triggering feedback. The AGN energy is supplied back to the ICM via bipolar precessing jets. \\
\indent
Compared to the feedback model used by \citet{YangReynolds2016b}, the main difference is that here we also include MHD and CR physics and consequently the energy injected by the AGN jets is supplied in kinetic and CR form. We consider jets dominated by the CR component and assume that a fraction of $f_{\rm cr}=0.9$ of the energy of the jet fluid is in the form of CRs. Other model parameters are: jet mass loading factor $\eta=1$, feedback efficiency $\epsilon=3\times 10^{-4}$, accretion timescale $t_{\rm ff}=5$ Myr, accretion radius $r_{\rm accre}=5.85$ kpc, precession period of the jet $t_{\rm prec}=$10 Myr, and precession angle of 15$^{o}$. The feedback energy is injected in a cylinder of 5 kpc in radius and 4 kpc in height. We refer the reader to \citet{YangReynolds2016b} and references provided therein for definitions of these quantities and further details. 
\subsection{Model equations}
We solve the MHD equations including CR advection, dynamical coupling between CR and the thermal gas, CR streaming along the magnetic field lines and the associated heating of the gas by CR, heating of the ICM by Coulomb and hadronic interactions, and radiative cooling
\begin{align}
\frac{\partial \rho}{\partial t} + \bnabla \cdot (\rho {\bf u}_{g})  &=  {\dot \rho}_{\rm j},\\
\frac{\partial \rho {\bf u}_{g}}{\partial t} + \nabla \cdot \left( \rho {\bf u}_{g}{\bf u}_{g}- \frac{{\bf B}{\bf B}}{4\pi} \right) + \bnabla p_{\rm tot} &=  \rho {\bf g} + {\dot p}_{\rm j},\\
\frac{\partial {\bf B}}{\partial t} - \bnabla \times ({\bf u}_{g}\times {\bf B})  & =  0, \label{eq:ind}\\
\frac{\partial e}{\partial t} + \bnabla \cdot \left[ (e+p_{\rm tot}){\bf u}_{g} - \frac{{\bf B}({\bf B} \cdot {\bf u}_{g})}{4\pi} \right]  &=  \rho {\bf u}_{g}\cdot {\bf g}\nonumber \\
-\nabla \cdot {\bf F}_{\rm c} - {\cal C} & +{\cal H}_{\rm c}+{\cal H}_{\rm j},   \label{eq:4} \\
\frac{\partial e_{\rm c}}{\partial t} + \bnabla \cdot (e_{\rm c} {\bf u}_{g}) = -p_{\rm c}\bnabla\cdot {\bf u}_{g}  -\nabla \cdot {\bf F}_{\rm c} & + {\cal C}_{\rm c}+{\cal H}_{\rm j},   \label{eq:5} 
\end{align}
where $\rho$ is the gas density, ${\bf u}_{g}$ is the gas velocity, ${\bf B}$ is the magnetic field, ${\bf g}$ is the gravitational field, ${\dot \rho}_{\rm j}$ is  the rate of injection of thermal gas via jet, ${\dot p}_{j}$ is the rate of momentum injection associated with the AGN, $e_{\rm c}$ is the specific CR energy density, and $e=0.5\rho {\bf u}_{g}^2 + e_{\rm g} + e_{\rm c} + B^2/8\pi$ is the total energy density, ${\cal C}$ is the radiative cooling energy loss rate per unit volume, ${\bf F}_{\rm c}$ is the CR flux due to streaming relative to the gas, ${\cal H}_{\rm c}$ is the rate of change of total specific energy due to streaming instability heating of the gas and Coulomb and hadronic CR losses, ${\cal C}_{\rm c}$ is the CR cooling rate due to the streaming instability, Coulomb, and hadronic CR losses, and ${\cal H}_{j}$ represents heating due to the AGN. The total pressure is $p_{\rm tot} = (\gamma_{g} -1)e_{\rm g} + (\gamma_{c} -1) e_{\rm c} + B^2/8\pi$, where $e_{\rm g}$ and $e_{\rm c}$ are the specific thermal energy density of the gas, $\gamma_{g}=5/3$ is the adiabatic index for ideal gas, and $\gamma_{c}=4/3$ is the effective adiabatic index of CR fluid. \\
\indent
Radiative cooling is included using the Sutherland \& Dopita cooling function \citep{SutherlandDopita93}. In order to speed up the computations we employ the sybcycling method \citep{anninos97, proga03} when the local cooling time becomes shorter than the hydrodynamical timestep.\\
\indent 
We solve the above equations using the adaptive mesh refinement MHD code FLASH4.2 \citep{fryxell2000,dubey2008}. We employ the directionally unsplit staggered mesh solver \citep{Lee09, Lee13}. This solver is based on a finite-volume, high-order Godunov scheme and utilizes a constrained transport method to enforce divergence-free magnetic fields.  We use third order MHD scheme and HLLD Riemann solver. 
\begin{table*}
  \caption{List of simulations}
  \label{tab:table2}
  \begin{center}
    \leavevmode
    \begin{tabular}{cccc} \hline \hline              
  run name &  Coulomb/hadronic heating & streaming heating & transport speed     \\ \hline 
  CHT0  & yes & no & 0  \\
  ST1  & no & yes & $v_{A}$ \\ 
  SCHT1  & yes & yes & $v_{A}$ \\ 
 SCHT4  & yes & yes & $4v_{A}$ \\ 
 CHT1  & yes & no &  $v_{A}$\\  \hline
  \multicolumn{4}{l}{}                           
   \end{tabular}\par
   \end{center}
\end{table*}
\subsection{Cosmic ray physics}
We include the heating of the ICM by CRs and transport of CRs with respect to the gas. Details of the CR physics module can be found in 
\citet{yang2012} and \citet{Ruszkowski2017}, where we discuss simulations the Fermi bubbles and  CR-driven galactic winds, respectively. We now summarize key CR physics processes described in that paper and discuss extensions of the CR module specific to the modeling of the ICM presented here.
\subsubsection{Streaming of cosmic rays}
Propagation of CRs in the magnetized ICM can be described in the framework of the self-confinement model. In this picture, CR scatter on waves excited by the streaming instability \citep{KulsrudPearce1969, Wentzel1974, Zweibel2013}. In a state of marginally stable anisotropy, the CRs stream at the Alfv{\'e}n speed down their pressure gradients.  However, the waves excited by the streaming instability can be damped by various mechanisms, e.g., by turbulent or Landau damping. When this happens, CRs can stream at speeds exceeding the Alfv{\'e}n speed. The effective streaming speed increases with the strength of the damping mechanism. 
The streaming flux is given by ${\bf F}_{\rm cr} = (e_{\rm cr} + p_{\rm cr}){\bf u}_{s}$, where ${\bf u}_{s} = -{\rm sgn}(\hat{{\bf b}}\cdot \nabla e_{cr})f{\bf u}_{A}$ is the streaming velocity, ${\bf u}_{A}$ is the Alfv{\'e}n velocity, and $f$ is the streaming speed boost factor.  \\
\indent
As demonstrated by \citet{wiener2013}, the effective streaming speed in the ICM can significantly exceed the Alfv{\'e}n speed in the cluster outskirts. For conditions representative of the cluster cool cores, damping mechanisms can lead to moderately super-Alfv{\'e}nic speeds for the following reasons. \citet{wiener2013} consider turbulent and non-linear Landau damping mechanisms. In the turbulent damping case, the effective streaming speed is 
\begin{equation}
u_{s}=u_{A}\left(1 + 0.08\frac{B_{10\mu G}^{1/2}n_{i,-2}^{1/2}}{L_{\rm mhd, 10}^{1/2}   n_{c,-9}}   \gamma_{3}^{n-3.5}10^{2(n-4.6)}\right),
\end{equation}
where $n_{i,-2}=n_{i}/(10^{-2}{\rm cm}^{-3})$ is the ion number density,  $n_{c,-9}=n_{c}/(10^{-9}{\rm cm}^{-3})$ is the CR number density, $L_{\rm mhd, 10}=L_{\rm mhd}/(10 {\rm kpc})$ is the lengthscale at which turbulence is driven at the Alfv{\'e}n speed $u_{A}$, $\gamma_{3}=\gamma/3$ is the average CR Lorentz factor, and $n>4$ is the slope of the CR distribution function in momentum (approximately $n=4.6$). 
In the non-linear Landau damping case the effective streaming speed is
\begin{equation}
u_{s}=u_{A}\left(1 + 0.03\frac{n_{i,-2}^{3/4}T_{5{\rm keV}}^{1/4}10^{n-4.6}\gamma_{3}^{(n-3)/2} } {B_{10\mu G}L_{\rm cr,10}^{1/2} n_{c,-9}^{1/2}}\right),
\end{equation}
where  $L_{\rm cr, 10}=L_{\rm cr}/(10 {\rm kpc})$ is the characteristic lengthscale of the fluctuations in the CR distribution and $T_{5{\rm keV}}=T/(5{\rm keV}$) is the ICM temperature. For the conditions representative of cool cores, in both of these cases, CR streaming is not typically super-Alfv{\'e}nic. However, the damping rate $\Gamma$ may be further boosted by linear Landau damping leading to $\Gamma_{\rm Landau}/\Gamma_{\rm turb}\sim \beta^{1/2}$, where $\beta$ is the plasma $\beta\sim 10^{2}$ parameter in the ICM (Zweibel, in prep.). When this process is included, the second term in Eq. (7) needs to be multiplied by $\beta^{1/2}$.  For plausible cool core parameters, the CR number density is  
\begin{equation}
n_{c}=3\times 10^{-9}\frac{n-4}{n-3}q_{-2}n_{i,-2}T_{5{\rm keV}}E_{\rm min, GeV}^{-1},
\end{equation}
where $q$ is the ratio of CR pressure to the ICM pressure and $E_{\rm min, GeV}$ is the low-energy cutoff in CR momentum distribution. Given the uncertainty in $\beta$, $L_{\rm mhd}$, and $n_{c}$, it is plausible that the effective CR streaming speed could be moderately super-Alfv{\'e}nic, i.e., boosted by a factor of order unity beyond the Alfv{\'e}n speed. Therefore, in addition to Alfv{\'e}nic streaming we also consider super-Alfv{\'e}nic streaming for $f=4$ in order to bracket our solutions.\\ 
\indent
CR streaming is incorporated using the method of \citet{sharma2009}. Because the term $-\nabla \cdot {\bf F}_{\rm cr}$ varies infinitely fast due to the discontinuity in the streaming flux near CR energy local extrema, it leads to a prohibitively small simulation timestep. In order to remove the singularity and speed up computations, we regularize the streaming flux by  ${\bf F}_{\rm c} = -(e_{\rm c}+p_{\rm c}){\bf u}_{A}{\rm tanh}(h_{\rm c}\hat{{\bf b}}\cdot\nabla e_{\rm c}/e_{\rm c})$, where $h_{c}$ is a free (regularization) parameter. In the calculations presented in this paper we adopt $h_{c}=100$ kpc.  
\subsubsection{ICM heating by cosmic rays}
As the CRs stream, they also experience an effective drag force. Consequently, CRs lose energy and the gas is heated due to the Alfv{\'e}n wave heating at the rate of
\begin{equation}
{\cal H}_{\rm cr, stream} = -{\bf u}_{A}\cdot\nabla p_{\rm cr}.
\end{equation}
In addition to the heating of the ICM associated with the streaming instability, CRs also heat the gas via Coulomb and hadronic interactions. We approximate the effects of CR cooling due to Coulomb and hadronic losses due to pion production via \citep{YoastHull2013} 
\begin{equation}
C_{{\rm cr}, c}=-4.93\times 10^{-19}\frac{n-4}{n-3}\frac{e_{c}\rho}{E_{\rm min}}\frac{\rho}{\mu_{e}m_{p}}\,{\rm erg\,}{\rm cm}^{-3}{\rm s}^{-1}
\end{equation}
and due to the hadronic losses via
\begin{equation}
C_{{\rm cr}, h}=-8.56\times 10^{-19}\frac{n-4}{n-3}\frac{e_{c}\rho}{E_{\rm min}}\frac{\rho}{\mu_{p}m_{p}}\,{\rm erg\,}{\rm cm}^{-3}{\rm s}^{-1}, 
\end{equation}
where $E_{\rm min}=1$ GeV is the minimum energy of CRs, $\mu_{e}$ and $\mu_{p}$ are the mean molecular weights per electron and proton, respectively. In the simulations we assume $n=4.5$ and mean proton Lorentz $\gamma = 3$. While all of the CR energy loss due to Coulomb collisions is transferred to the gas, only $\sim1/6$ of the CR energy loss due to pion production is used to heat the gas and the remainder is removed as gamma ray emission and neutrinos. Consequently, the rate of change of the total specific energy density of the gas, that includes the thermal and CR specific energy densities, is ${\cal H}_{\rm cr}=(5/6)C_{{\rm cr},h}/\rho<0$ and the CR specific energy density loss rate is ${\cal C}_{\rm c}=(C_{{\rm cr},c}+C_{{\rm cr},h})/\rho$. 
\section{Results}
\begin{figure*}
  \begin{center}
    \leavevmode
        \includegraphics[width=0.24\textwidth]{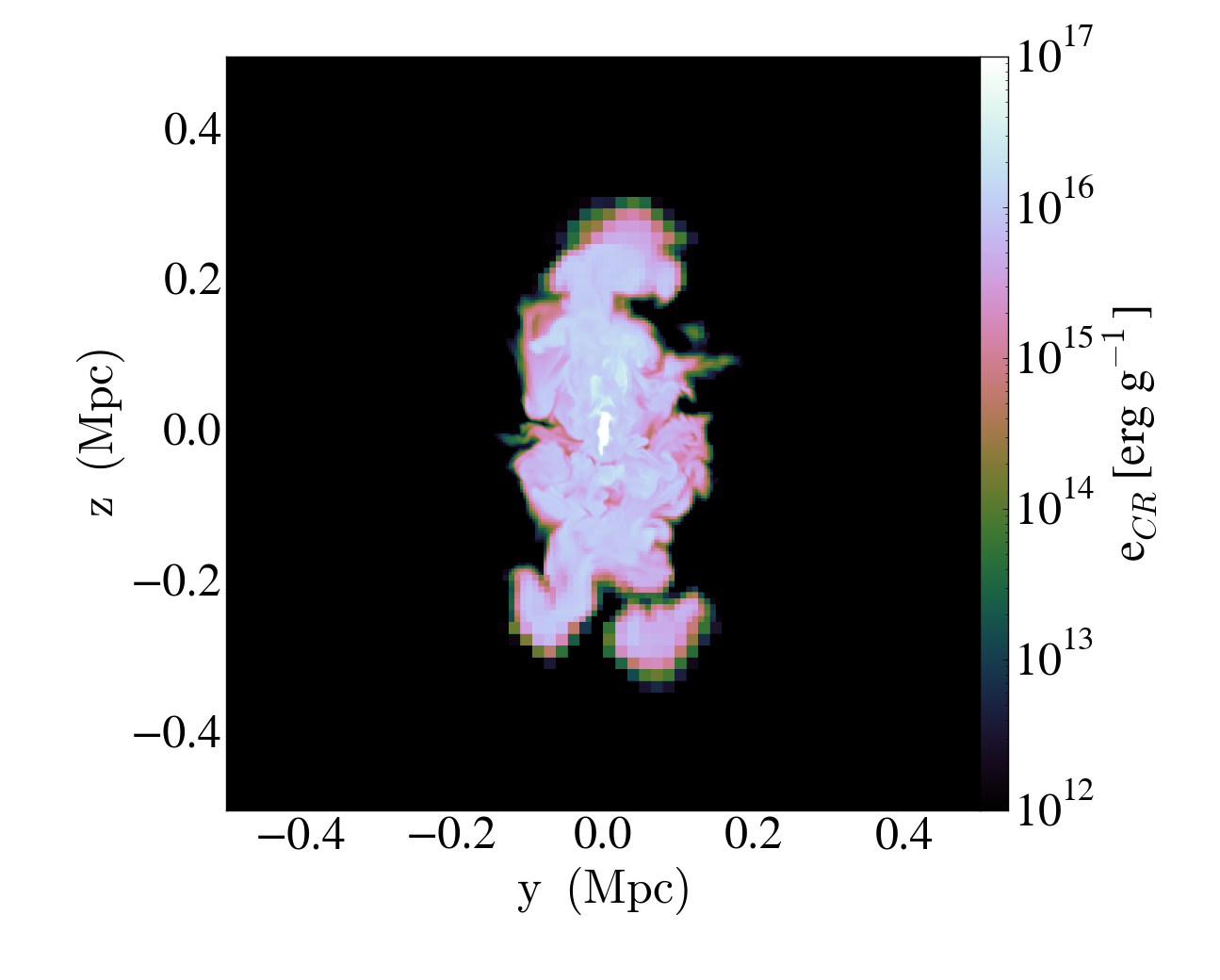}
        \includegraphics[width=0.24\textwidth]{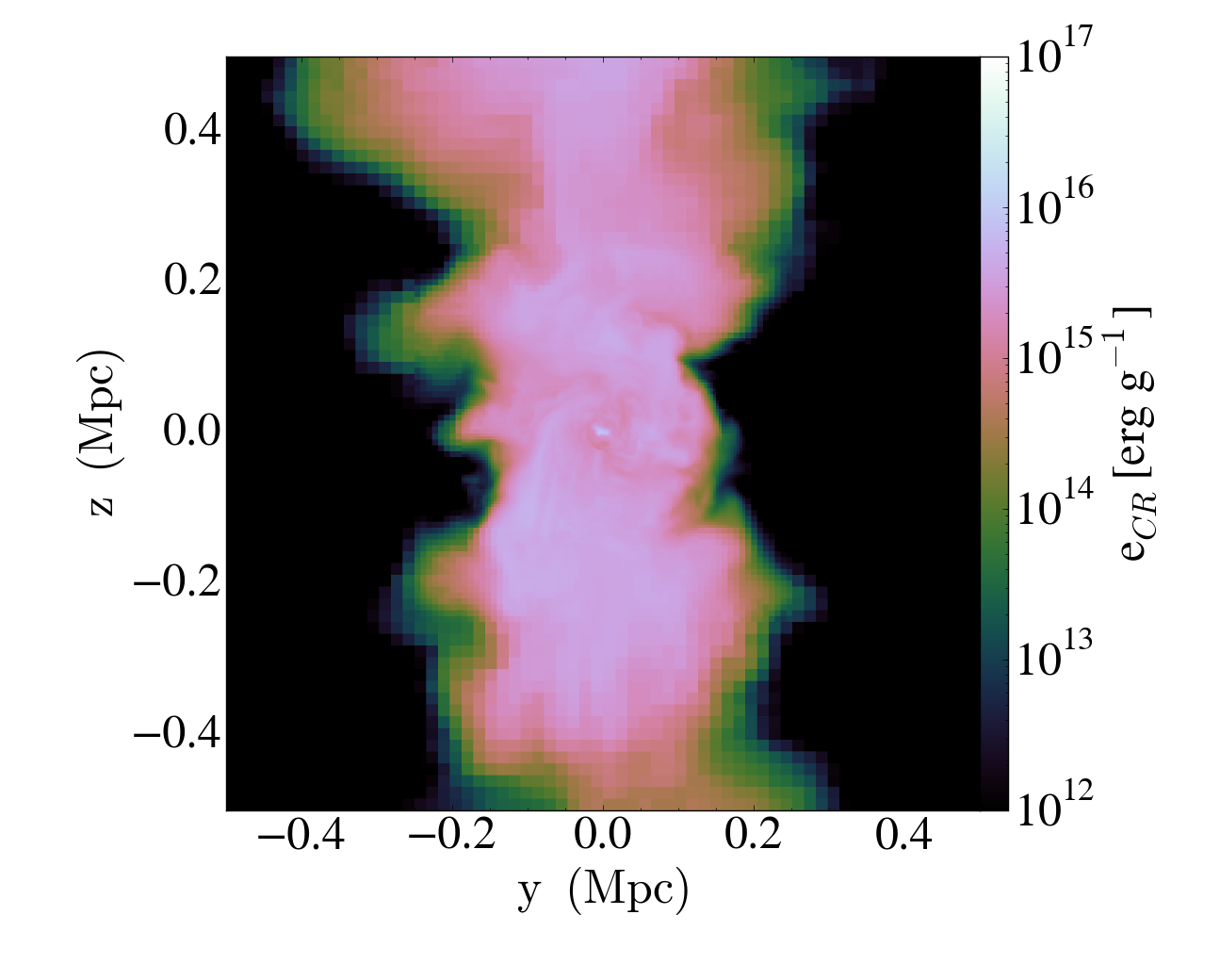} 
        \includegraphics[width=0.24\textwidth]{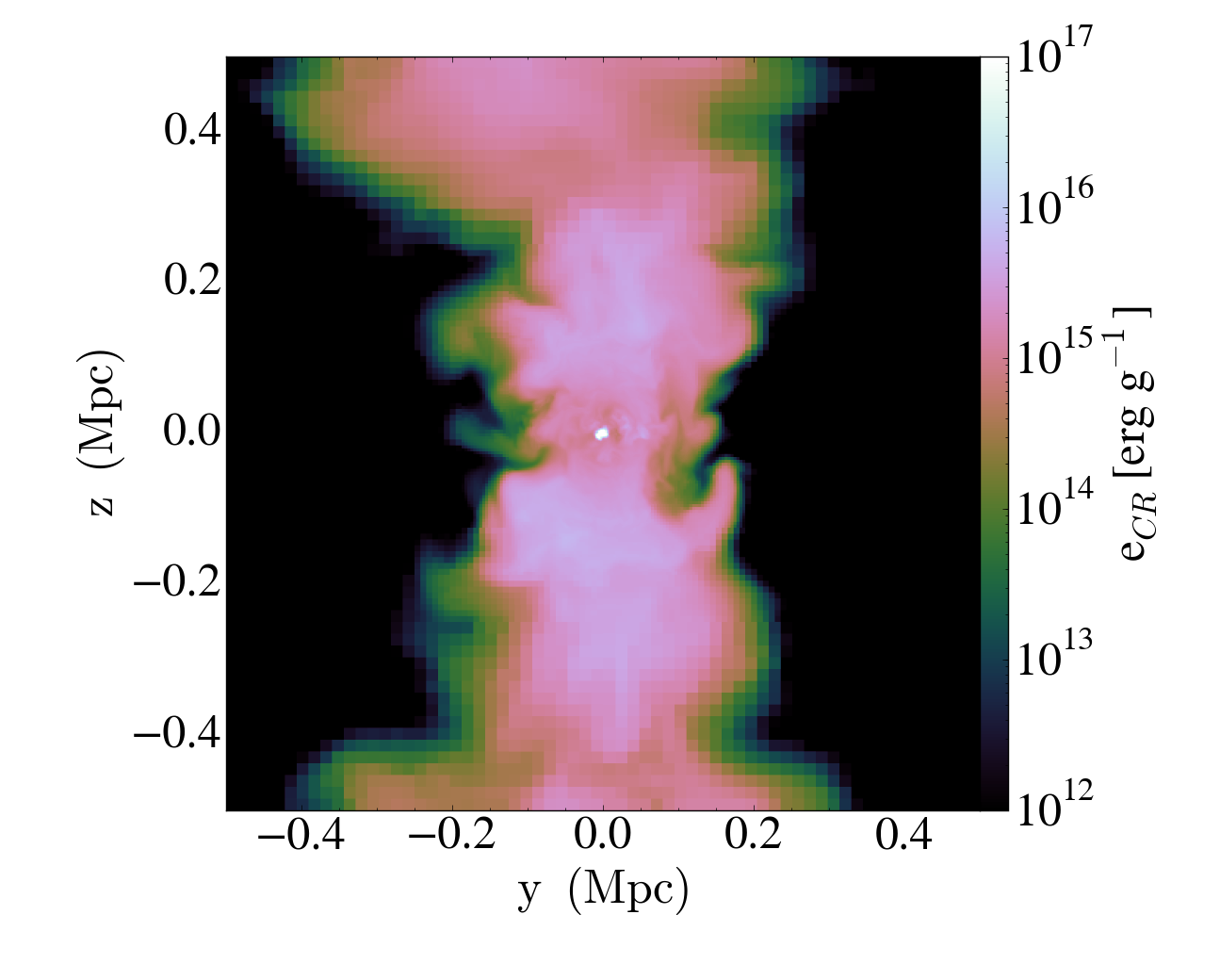}   
        \includegraphics[width=0.24\textwidth]{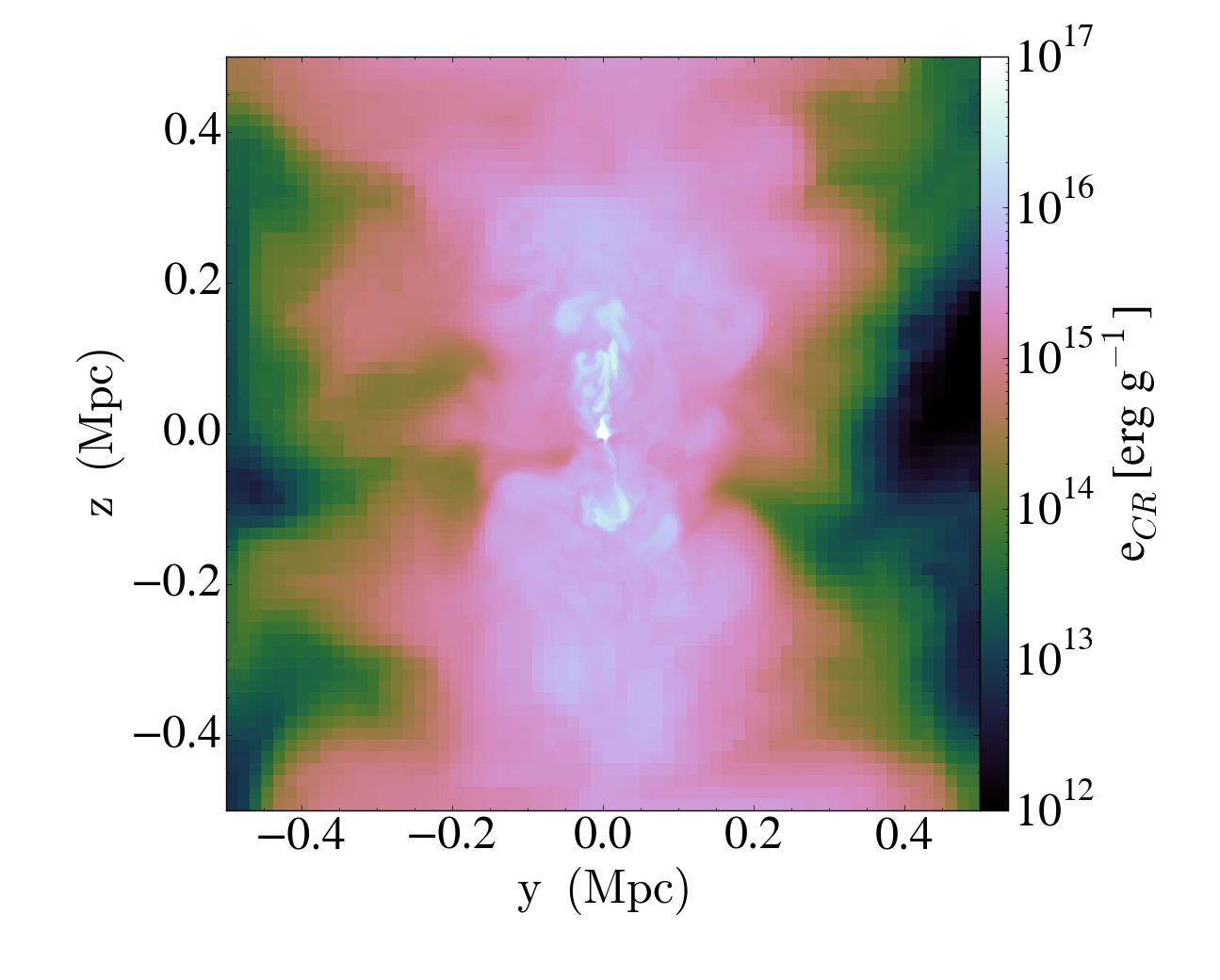}            
       \caption[]{From left to right: Slice though the cosmic ray energy density distribution for the case with hadronic and Coulomb heating (CHT0), 
    cosmic ray streaming/heating (ST1), cosmic ray streaming/heating and hadronic and Coulomb heating (SCHT1), and same as the last panel but for super-Alfv{\'e}nic streaming (SCHT4). All snapshots were taken at 3 Gyr.}
     \label{fig:dens}
  \end{center}
\end{figure*}
\indent
The list of the performed runs is shown in Table 1.  Figure 1 presents cross sections through the cluster center showing the distribution of the specific CR energy density. From left to right these slices correspond to the following cases: (i) hadronic and Coulomb heating but no transport processes (CHT0), (ii) CR streaming and streaming heating (ST1), (iii) CR streaming and heating due to streaming, hadronic and Coulomb processes (SCHT1), and (iv) same as the last panel but for super-Alfv{\'e}nic streaming (SCHT4). All snapshots were taken at 3 Gyr. This figure demonstrates that CR transport processes affect the morphology of the radio emitting plasma and effectively redistribute CRs. The redistribution of CR energy is efficient despite the fact the jet is pointed in approximately constant direction. As expected, the widening of the CR distribution is most significant when the CR transport is the fastest, i.e., super-Alfv{\'e}nic. Note that these results also imply that the dynamical state of the atmosphere does depend on whether CR transport is included. Despite the fact that all snapshots were taken at the same time, the case where the CR streaming is neglected corresponds to the most perturbed atmosphere at the center of the cool core, while in all cases that include streaming, the ICM is relatively less disturbed and calmer at this particular time. As described in detail below, in the simulations including CR streaming the ICM generally exhibits larger variations due to more intermittent AGN feedback. This means that the atmosphere can experience both the periods of relative calm and more stormy conditions. Recent Perseus data from Hitomi is consistent with relatively low level of turbulence in this cluster \citep{Hitomi2016}. It is plausible that the dynamical state of the Perseus cluster currently corresponds to relatively low-turbulence state captured in Figure 1 in cases including transport processes (see also \citet{Li2016}).  Alternatively, turbulent motions in the cluster atmosphere could be reduced due to viscosity. We also point out that the iron line shifts corresponding to large gas velocities induced by the AGN at the center of the cool core may be partially diluted by slower moving gas away from the center. This may give an impression of relative calm in the ICM even if fast gas motions are present. This dilution effect has been seen in mock Hitomi simulations that show line shifts consistent with the data (Morsony, priv. comm.). We defer to a future publication the study of the iron emission line profiles and observational predictions for the planned Hitomi replacement and the X-ray Surveyor missions. \\
\indent
As expected, the dispersal of CRs throughout the core is more pronounced at later times since the onset of feedback and when the speed of CR transport is faster. Interestingly, observations of M87 with LOFAR reveal a sharp radio emission boundary that does not seem to depend sensitively on radio frequency \citep{deGasperin2012}, i.e., it appears that the boundary corresponds to the physical extent of CRs. At late times no such boundary is seen in the simulations. However, such boundary in the spatial distribution of CRs could be explained by large-scale sloshing motions that order magnetic fields on large scales and prevent the leakage of CRs to large distances by suppressing cross-field CR transport. Simulations of \citet{zuhone2013} show that sloshing motions induced by substructure in the cluster can generate tangential magnetic fields. Such fields could slow down radial transport of CRs away from the core. Alternatively, weaker or less collimated AGN feedback could prevent the bubbles from overshooting the critical radius at which their internal entropy equals that of the ambient ICM. In such a case, we would expect CR to exist predominantly within such critical radius. We defer exploration of these possibilities to a future publication and point out that there exist counter-examples to  the morphological appearance of M87. In Abell 262 \citep{clarke2009} and A2597 \citep{clarke2005} the radio emission at lower frequencies extends to larger distances from the cluster center.\\
\begin{figure*}
  \begin{center}
    \leavevmode
        \includegraphics[width=0.24\textwidth]{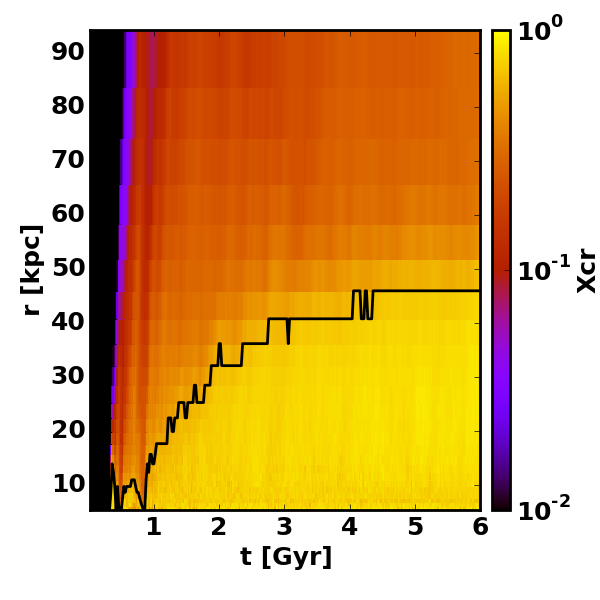}
        \includegraphics[width=0.24\textwidth]{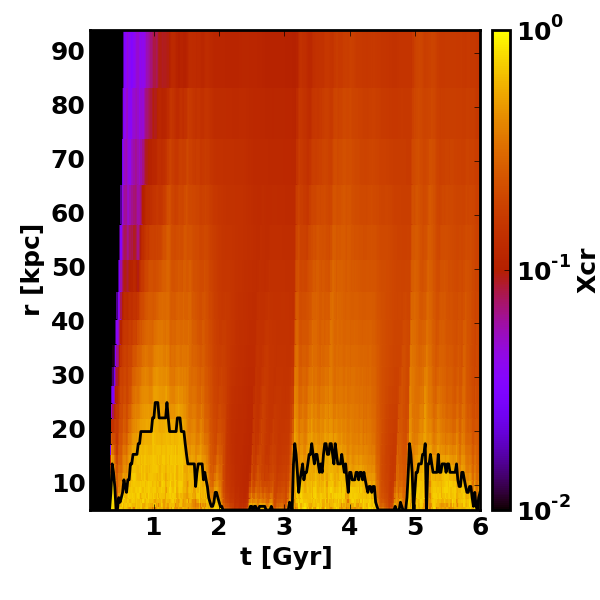} 
        \includegraphics[width=0.24\textwidth]{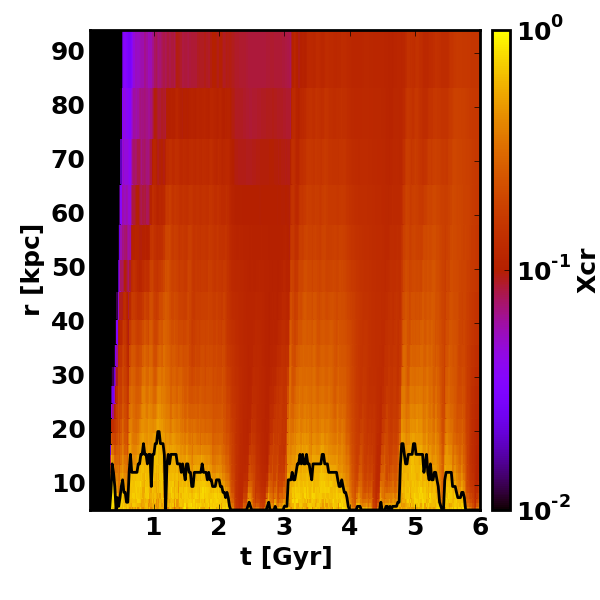}    
        \includegraphics[width=0.24\textwidth]{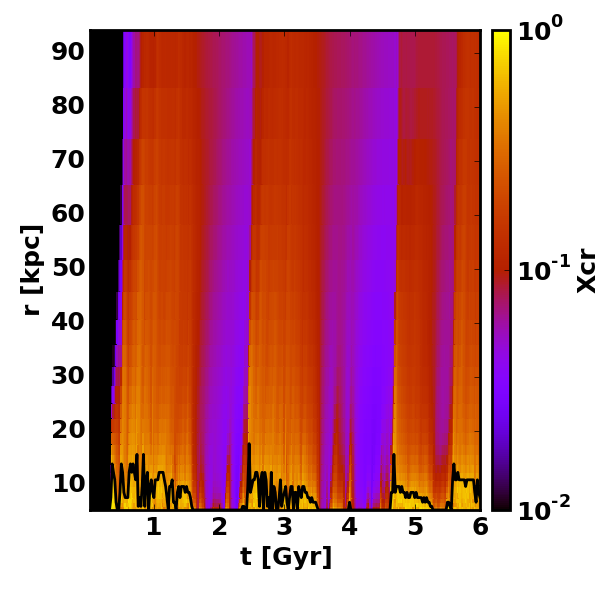}                      
       \caption[]{Evolution of cosmic ray pressure support distribution in the intracluster medium (ordering of  panels is the same as in Fig. 1). Dark line corresponds to 50\% contribution to pressure support.}
     \label{fig:dens}
  \end{center}
\end{figure*}
\indent
The pressure support due to CRs is quantified in Figure 2. Pressure support is defined as the ratio of the pressure provided by CRs to the sum of the thermal and CR pressures. In order to exclude CR-filled bubbles that are cooling very inefficiently, this quantity is set to $10^{-2}$ whenever the local cooling time exceeds the Hubble time. All panels show the evolution of the profiles of the pressure support. Dark lines corresponds to 50\% of CR contribution to the total pressure support. In the case excluding CR transport (left panel), CR interaction with the ambient medium is inhibited. This is caused by the presence of the magnetic fields that slow down the mixing process and the fact that CRs are simply advected with the gas and do not stream with respect to the location of the fluid injected by the AGN. Consequently, even though hadronic and Coulomb heating processes are included, the CR heating of the ambient ICM is ineffective because CRs do not easily come in contact with the thermal ICM. This means that the cooling catastrophe can easily develop, which leads to large mass accretion rates onto the central supermassive black hole. As a result of this accretion the black hole feedback increases and more CRs are injected into the ICM. This is a runaway process in which CRs account for progressively larger fraction of the total pressure support. At the end of the simulation the CR pressure support in $\sim$50 kpc is dominant and thus it is inconsistent with observational constraints \citep{JacobPfrommer2016b,JacobPfrommer2016b}.\\
\indent
The remaining three panels illustrate that the role of transport processes is essential for removing this tension with observations. The second panel shows that including CR streaming and associated with it streaming heating dramatically reduces CR contribution to the pressure support. This reduction in CR pressure occurs because CRs can now come into contact with the thermal ICM and heat it, thus reducing the CR energy density and associated with it CR pressure. Similarly, CR pressures are further reduced when, in addition to the processes included in the second panel, we also include CR hadronic and Coulomb losses. These two processes further drain the energy from CRs and heat the thermal gas. Finally, the last panel demonstrates the consequences of including faster (super-Alfv{\'e}nic) streaming. As expected, this further reduces CR pressure support. Note that this boost in the CR streaming speed only affects the rate of CR transport rather than the Alfv{\'e}n wave heating. In all cases but the one shown in the leftmost panel, the CR pressure support is very small. \\
\indent
We also performed a run without streaming instability heating but including transport by streaming and heating by Coulomb and hadronic processes (CHT1; not shown). While this run is unphysical, it helps us to better understand the role of CR transport. In this run, the values of CR pressure support (and typical variability timescales of CR pressure support; see below for more detailed discussion of variability) are similar to those seen in the three right panels in Figure 2. This experiment shows that CR transport is essential for preventing cooling cathastrope.\\
\begin{figure*}
  \begin{center}
    \leavevmode
       \includegraphics[width=0.24\textwidth]{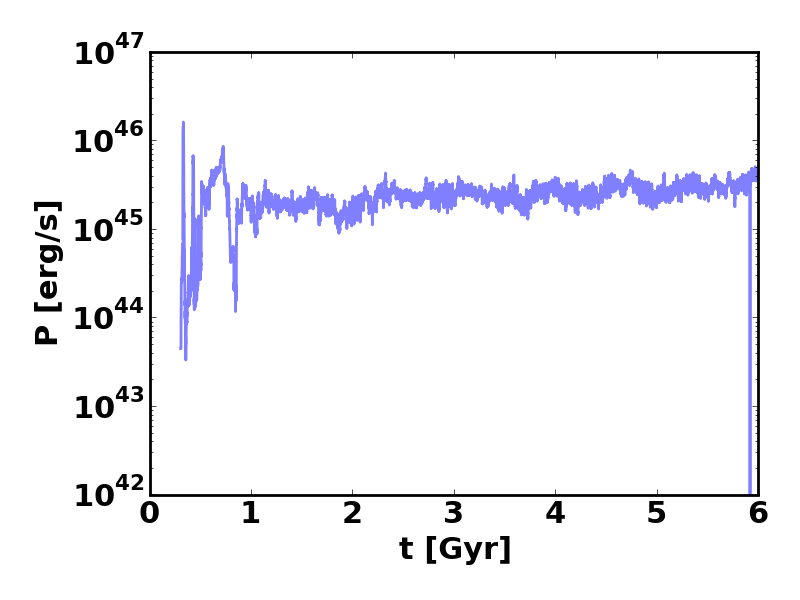}
        \includegraphics[width=0.24\textwidth]{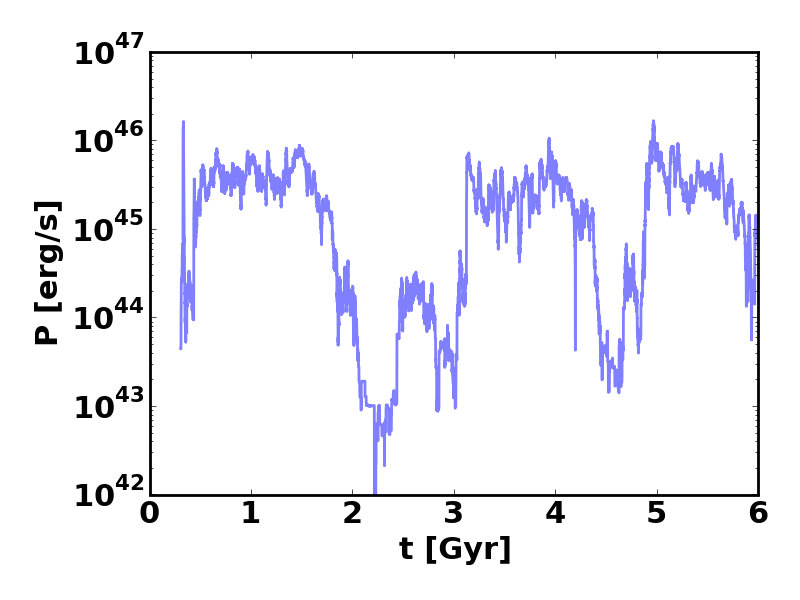} 
        \includegraphics[width=0.24\textwidth]{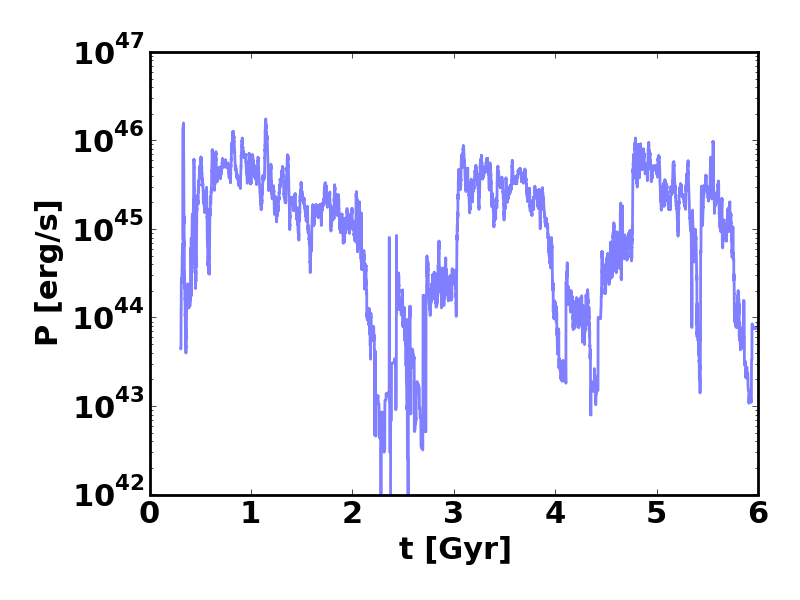}   
        \includegraphics[width=0.24\textwidth]{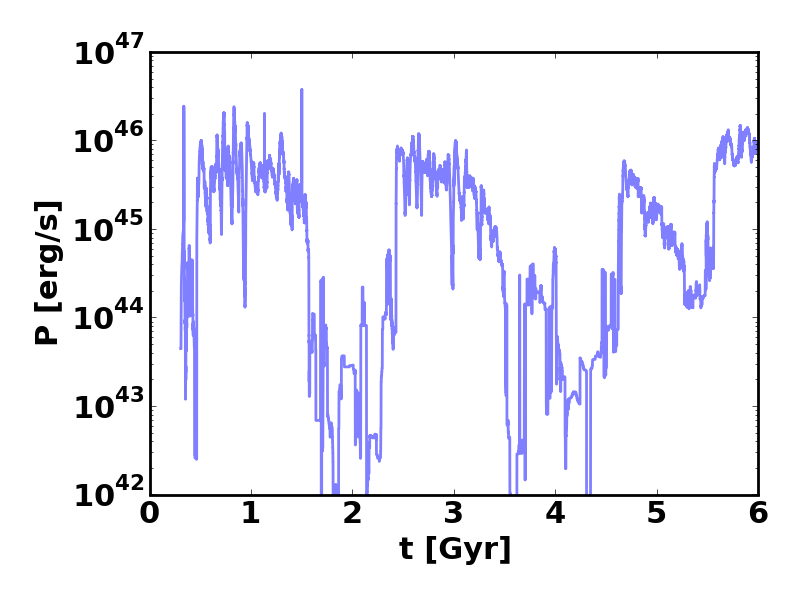}                 
       \caption[]{Jet power (ordering of figures is the same as in Fig. 1). }
     \label{fig:dens}
  \end{center}
\end{figure*}
In all three cases that include transport processes (panels 2 to 4 in Figure 2) there is a significant variation in the CR pressure support over time. This is a consequence of self regulation of the AGN feedback that was absent from the non-streaming case (panel 1 in Figure 2) where a global runaway cooling instability dominated the evolution of the ICM. This self-regulating behavior of the atmosphere is reflected in Figure 3
which shows AGN jet power as a function of time. In all four cases but the one shown in the first panel, the black hole feedback is highly variable. Note that despite the large variability, the AGN jet never completely switches off. \\
\indent
While predicting detailed observational gamma-ray and radio signatures based on these simulations is beyond the scope of this paper, we point out that 
typical levels of CR pressure support that we find in simulations including CR transport are generally consistent with the data. Based on one-dimensional models that include heating by thermal conduction and CRs,  \citet{JacobPfrommer2016b} argue that in those cool core clusters that do not host radio mini halos, AGN activity and CR heating are the strongest, and that CRs can provide adequate level of heating without violating observational radio and gamma-ray constraints. They further argue that primary and secondary CR electron radio emission associated with the AGN outbursts could be difficult to detect due to the small physical extent of the radio emission in this case and the large flux dynamical range of the AGN jet and the halo. This picture is likely to be consistent with the elevated CR pressure support during AGN outbursts that is seen in Figure 2 (e.g., near $\sim3$ Gyr in the rightmost panel). In \citet{JacobPfrommer2016a} typical values of CR-to-thermal pressure are on the order of 0.1 and vary substantially from object to object and thus presumably depend on the cluster dynamical state. Interestingly, \citet{Pfrommer2013} shows that in the Virgo cluster, in the absence of thermal conduction, adequate CR heating rate can be supplied when CR fraction is around 0.3 while not violating observational data. Levels of CR pressure support that we observe in our simulations during outbursts are comparable to those suggested by \citet{JacobPfrommer2016a} and could presumably be reduced further if we included thermal conduction.  In the case of cool cores that are associated with radio mini halos,  \citet{JacobPfrommer2016b} predict that the amount of CR pressure support needed to stably heat the cool core exceeds observational limits and suggest that such objects are expected to be dominated by radiative cooling. This situation could correspond to the periods in between the outbursts seen in Figure 2. Thus, the general properties of our simulations, and in particular the presence of the feedback loop and two classes of cool cores, are broadly consistent with the picture based on the above one-dimensional models. We also note that the simulations that do not include CR transport processes (left panel in Figure 2), do not show intermittent AGN activity and would therefore not be able to account for the transitions between cool cores with and without radio mini halos. Finally, we note that here we focus on general trends and defer to future publication the study of the parameter space of the models that meet observational constraints in detail. \\
\begin{figure*}
  \begin{center}
    \leavevmode
       \includegraphics[width=0.24\textwidth]{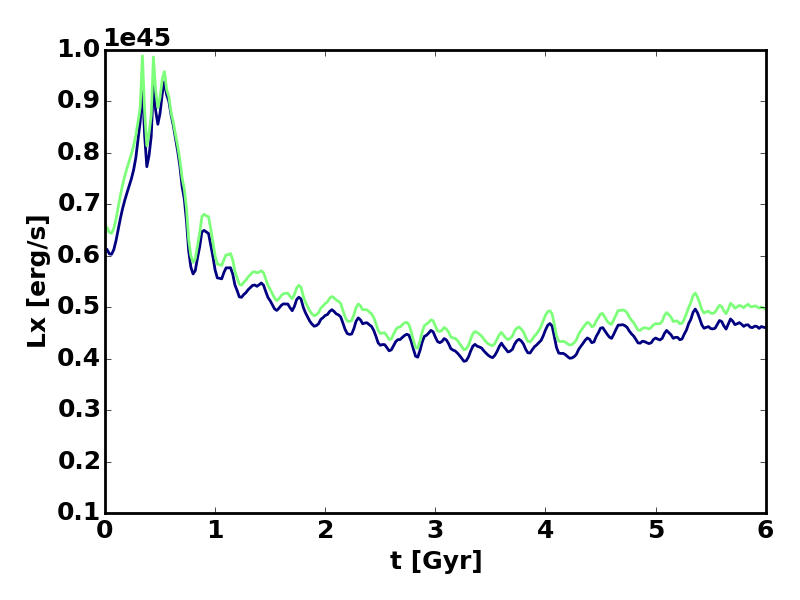}
        \includegraphics[width=0.24\textwidth]{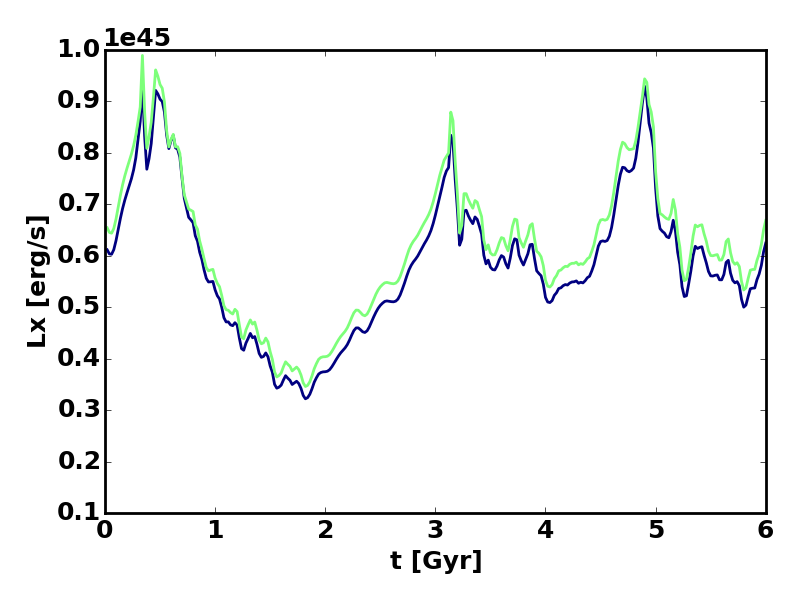} 
        \includegraphics[width=0.24\textwidth]{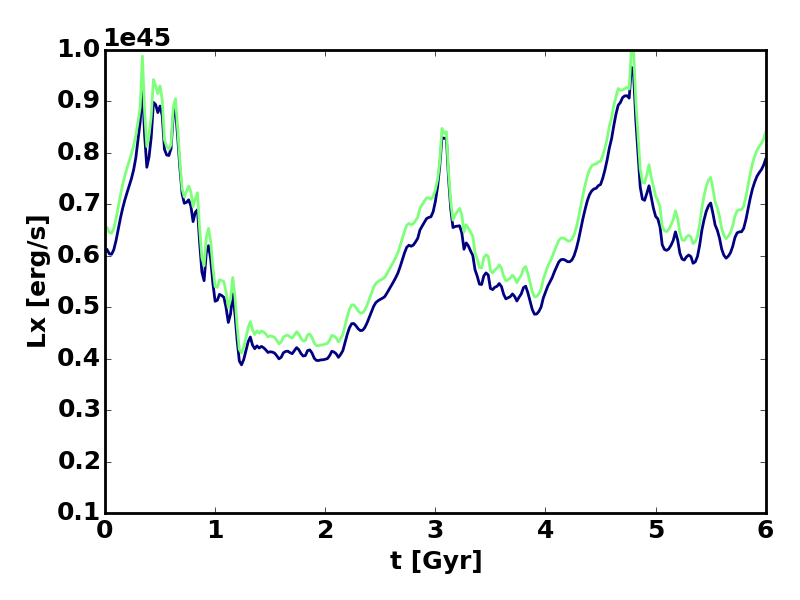}      
        \includegraphics[width=0.24\textwidth]{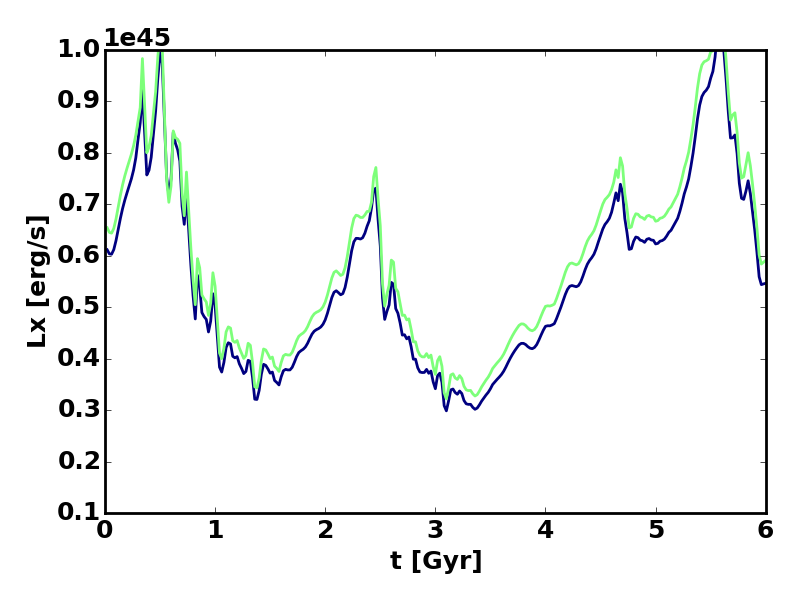}                 
       \caption[]{X-ray luminosity within the central 100 kpc (ordering of panels is the same as in Fig. 1). Green line corresponds to bolometric brehmsstrahlung luminosity and black line to the X-ray emission integrated in the 0.5 to 10 keV range.}
     \label{fig:dens}
  \end{center}
\end{figure*}
The evolution of the X-ray luminosity within the central 100 kpc is shown in Figure 4. Green line corresponds to bolometric brehmsstrahlung luminosity and black line to the X-ray emission integrated in the 0.5 to 10 keV range. As the X-ray emission is dominated by the densest central region of the cool core, an increase in the X-ray luminosity implies larger accretion of gas onto the central supermassive black hole. This boost in the accretion rate consequently implies stronger AGN feedback and this is why peaks in the X-ray luminosity closely correlate with the times when the jet power increases (see Figure 3). This cyclic behavior of the X-ray luminosity is evident in the cases including CR streaming.\\
\begin{figure*}
  \begin{center}
    \leavevmode
         \includegraphics[width=0.33\textwidth]{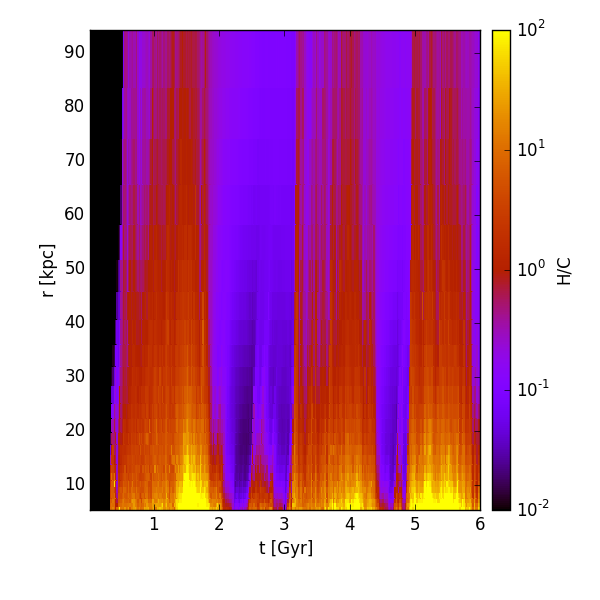}
         \includegraphics[width=0.33\textwidth]{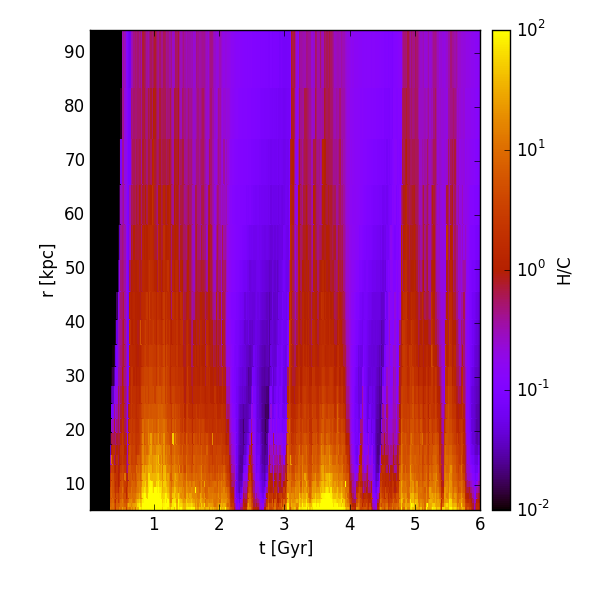} 
          \includegraphics[width=0.33\textwidth]{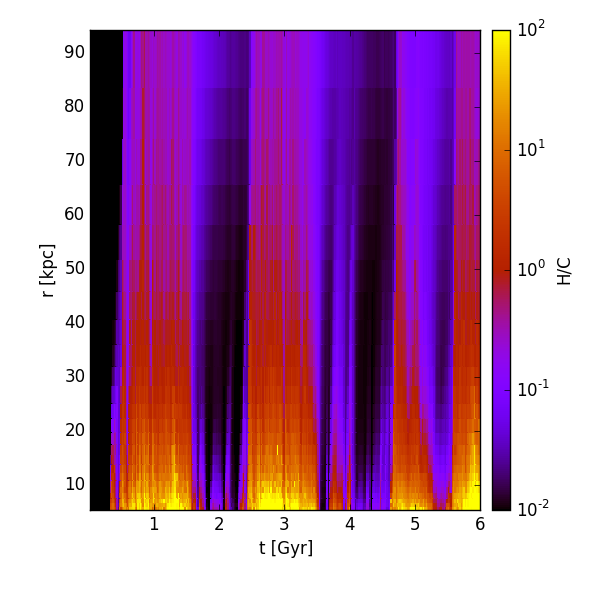}         
         \includegraphics[width=0.33\textwidth]{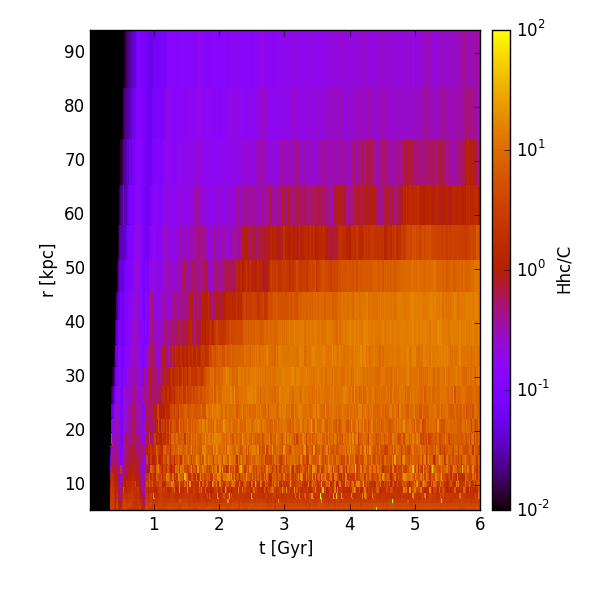}     
         \includegraphics[width=0.33\textwidth]{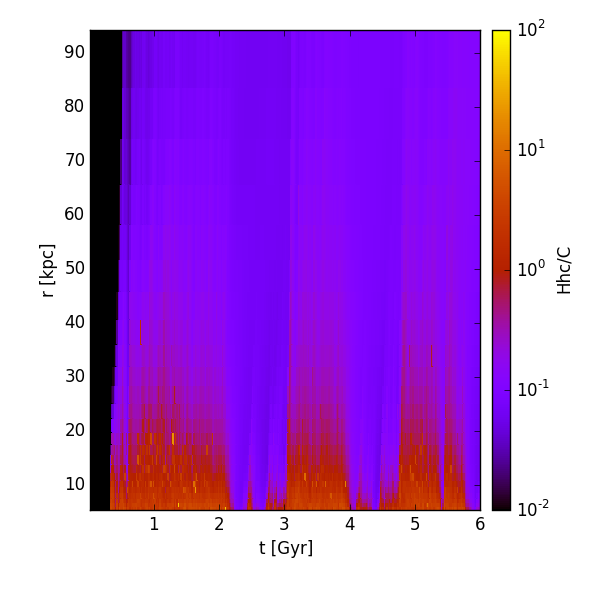}     
         \includegraphics[width=0.33\textwidth]{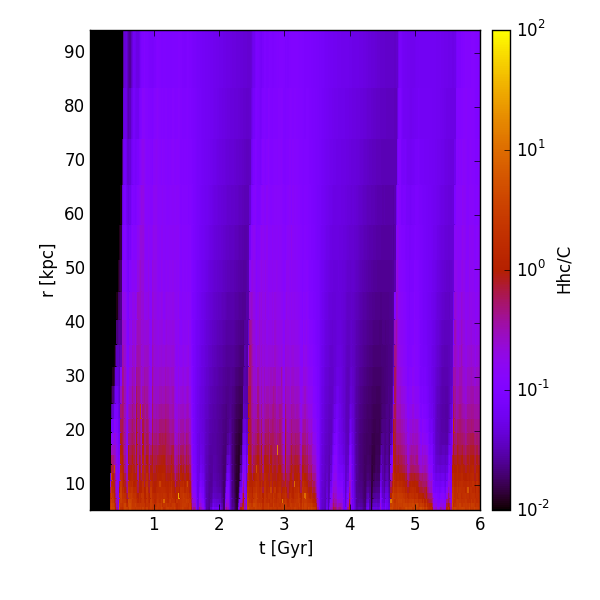}          
       \caption[]{Evolution of the distribution of the ratio of cosmic ray heating to radiative cooling in the intracluster medium. From left to right, top row corresponds to the heating due to streaming in the case with: (i) streaming heating (ST1), (ii) streaming heating and hadronic and Coulomb heating (SCHT1), (iii) super-Alfv{\'e}nic streaming heating and hadronic and Coulomb heating (SCHT4). Bottom row shows the ratio of combined Coulomb and hadronic heating to radiative cooling. Shown from left to right in the bottom row are the following cases: (i) Coulomb and hadronic heating without CR streaming transport (CHT0), (ii) Coulomb and hadronic heating with CR streaming transport (SCHT1), (iii) same as (ii) but for super-Alfv{\'e}nic CR streaming transport (SCHT4).}     \label{fig:dens}
  \end{center}
\end{figure*}
The evolution of the profiles of the ratio of heating to radiative cooling is shown in Figure 5. As in the case of the profiles of the CR pressure support shown in Figure 2,  in order to exclude regions that are cooling very inefficiently, the heating-to-cooling ratio is set to $10^{-2}$ whenever the local cooling time exceeds the Hubble time. From left to right, top row corresponds to the heating due to streaming in the case with: (i) streaming heating (ST1), (ii) streaming heating and hadronic and Coulomb heating (SCHT1), (iii) super-Alfv{\'e}nic streaming heating and hadronic and Coulomb heating (SCHT4). Bottom row shows the ratio of combined Coulomb and hadronic heating to radiative cooling. Shown from left to right in the bottom row are the following cases: (i) Coulomb and hadronic heating without CR streaming transport (CHT0), (ii) Coulomb and hadronic heating with CR streaming transport (SCHT1), (iii) same as (ii) but for super-Alfv{\'e}nic CR streaming transport (SCHT4).\\
\indent
Let us begin discussing Figure 5 by focusing on the bottom left panel. This panel shows the ratio of the combined heating due to Coulomb and hadronic interactions to radiative cooling without including CR transport effects. This panel mirrors what is shown in the left panel in Figure 2 in the sense that the regions characterized by high heating-to-cooling ratios increase in size over time just as the regions occupied by high CR fraction grow with time. This significant heating is a direct consequence of the accumulation of large amounts of CRs in the cluster center. The accumulation of CRs is caused by increased AGN energy injection. However, because the mixing of CRs with the ICM is inefficient in this case, bulk of the ICM begins to overcool, which in turn leads to the stronger AGN feedback and associated with it CR injection. This particular case is ultimately unsuccessful because the CR heating does not couple well to the bulk of the ambient ICM. This is also consistent with the evolution of the jet power shown in Figure 3. By comparing the leftmost panel in Figure 3 that corresponds to the case without CR streaming to the jet power evolution in the cases that do include streaming (panels 2 through 4 in Figure 3), one can see that the integrated jet power, and thus the amount of CRs that accumulate in the cluster core, is the largest in the non-streaming case. When CR transport is neglected, the coupling of CRs to the gas is very weak. Consequently, gas accretion is unopposed, jet is constantly turned on, but its energy is not used efficiently to offset radiative losses in the ICM. Thus, accretion proceeds uninterrupted, and the AGN is not intermittent.\\
\indent
The non-streaming case is deceptively similar to the cases considered by \citet{YangReynolds2016b}, who simulated AGN feedback using hydrodynamical simulations. 
The main differences between the non-streaming case presented here and their simulations is that (i) in their model AGN jets inflate bubbles dominated by thermal energy whereas in our case the injection is dominated by CRs, and (ii) we include magnetic fields. Even though hadronic and Coulomb interactions are included in the non-streaming case, mixing of the AGN fluid with the ambient ICM is inhibited by magnetic fields and so the coupling of the AGN fluid to the ambient thermal gas is suppressed. This suppression is absent from \citet{YangReynolds2016b} simulations and the heating of the ambient ICM can proceed via mixing with the thermal AGN jet fluid. This interpretation is also consistent with the results of \citet{sijacki2008} who do include CRs but neglect magnetic fields. In their case cooling catastrophe is prevented most likely as a result of more efficient mixing of the AGN fluid containing CRs and subsequent interactions of CRs with the ambient ICM via processes other than streaming heating.\\
\indent
The fact that the non-streaming case fails to self-regulate also implies that other heating mechanisms such as dissipation of turbulence or weak shocks, though present, are not the dominant sources of heating of the ICM. Instead, CR heating through interaction between the CRs and the ICM is essential for reaching a global thermal balance. This conclusion is analogous to that of \citet{YangReynolds2016a} who point out a similar hierarchy of heating sources, but that the role of CR heating in our simulations is replaced by mixing of the ultra-hot gas within the bubbles with the ambient ICM in the hydrodynamic case.  \\
\begin{figure*}
  \begin{center}
    \leavevmode
        \includegraphics[width=0.24\textwidth]{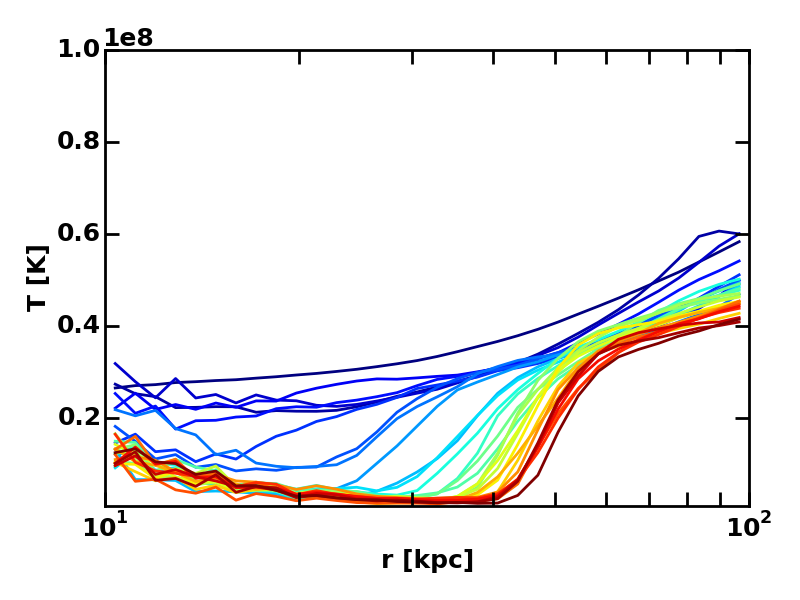}
        \includegraphics[width=0.24\textwidth]{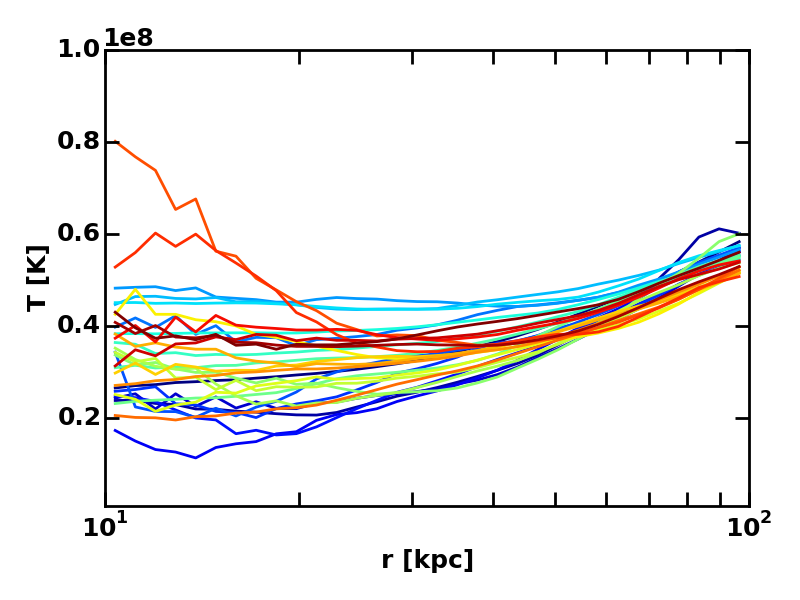} 
        \includegraphics[width=0.24\textwidth]{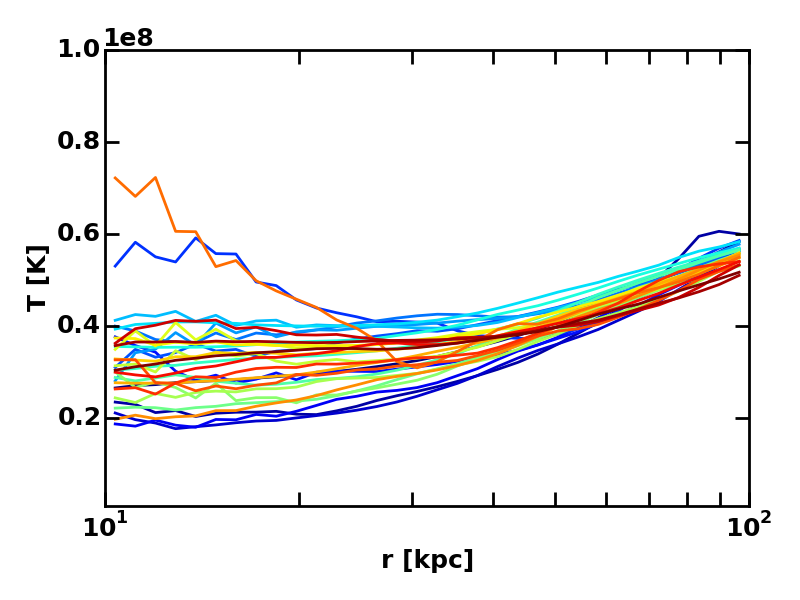}        
        \includegraphics[width=0.24\textwidth]{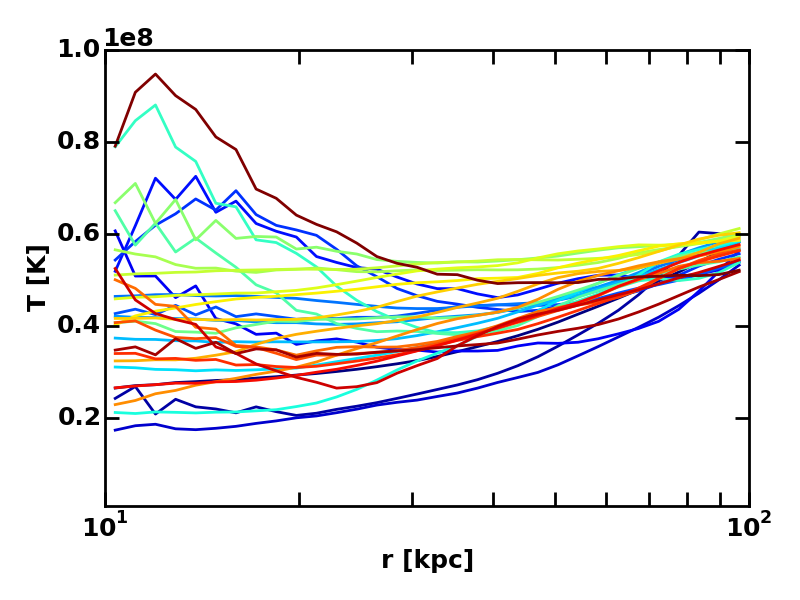}        

        \includegraphics[width=0.24\textwidth]{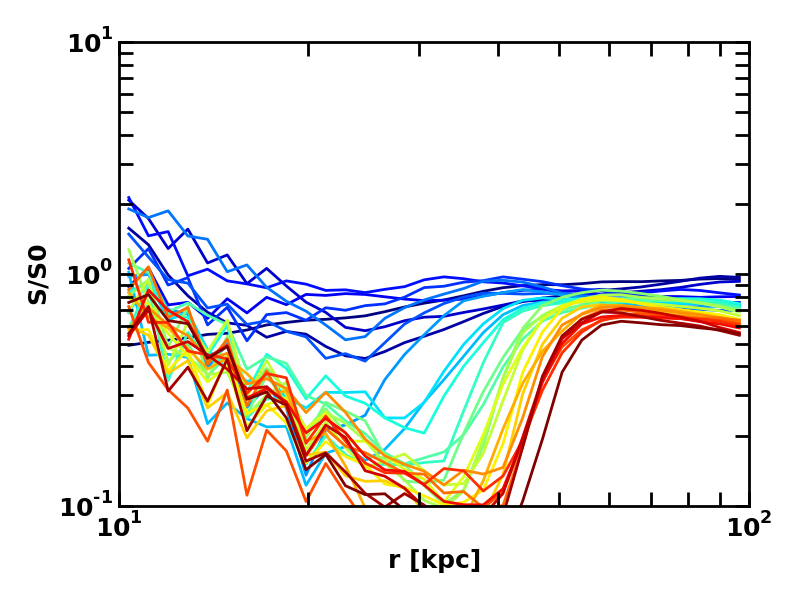}
        \includegraphics[width=0.24\textwidth]{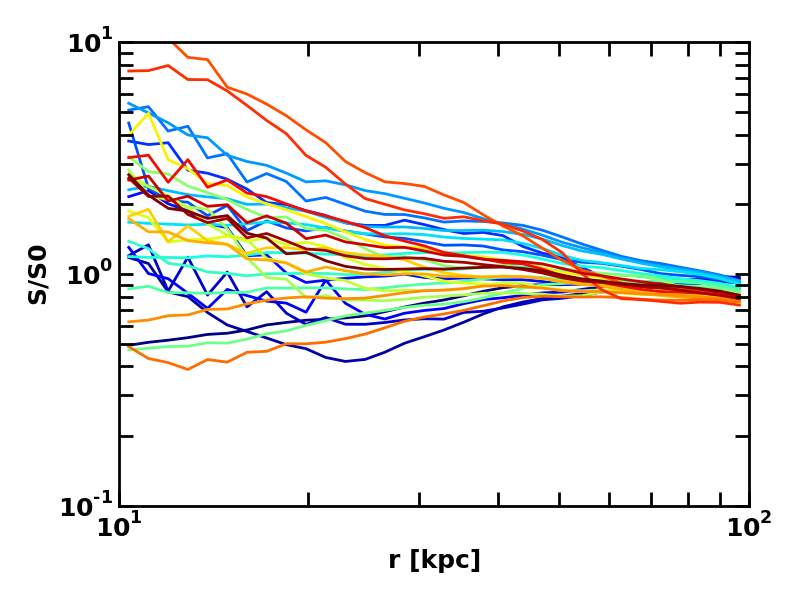} 
        \includegraphics[width=0.24\textwidth]{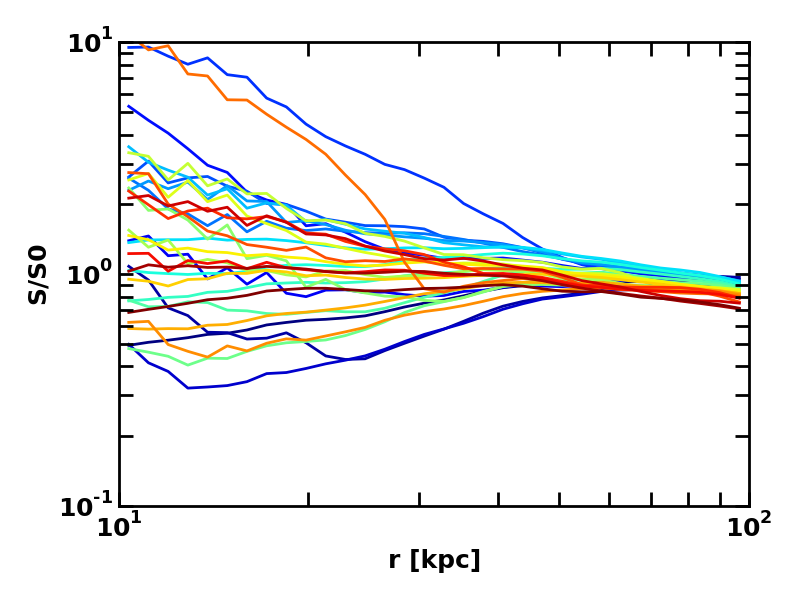}     
        \includegraphics[width=0.24\textwidth]{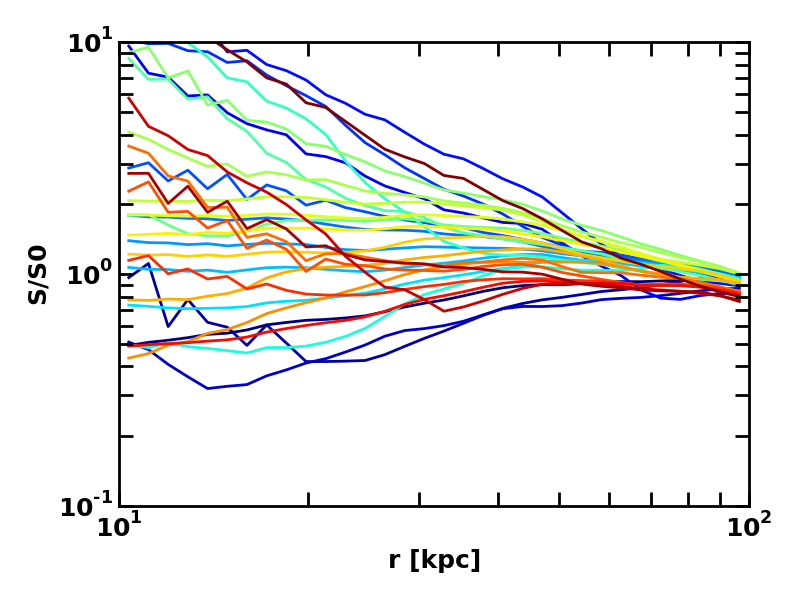}    
 
        \includegraphics[width=0.24\textwidth]{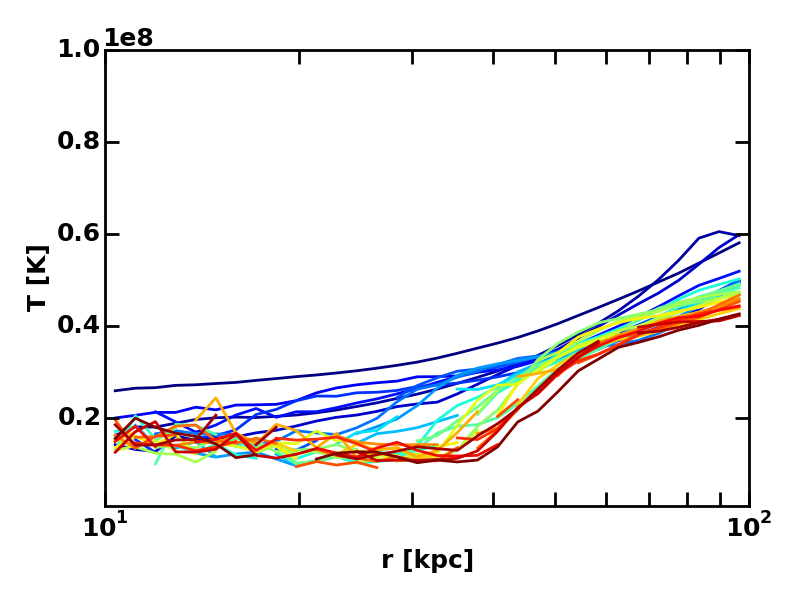}
        \includegraphics[width=0.24\textwidth]{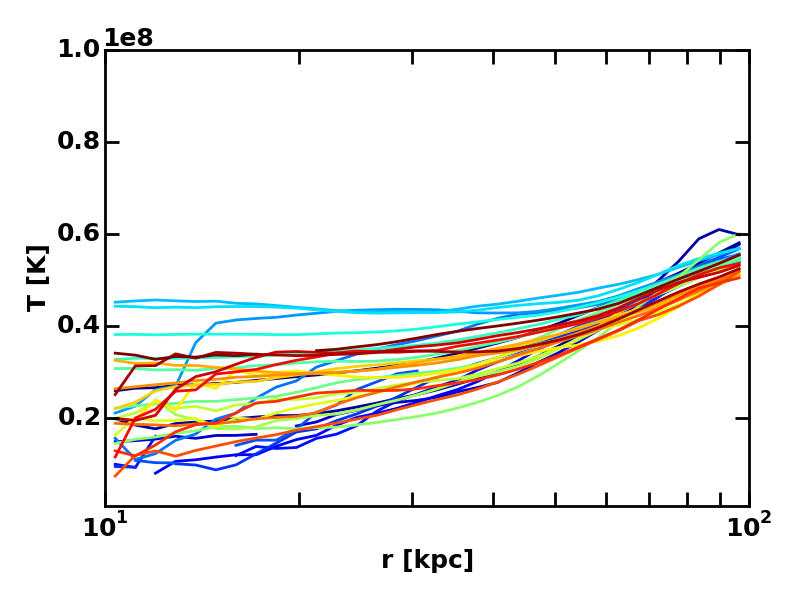} 
        \includegraphics[width=0.24\textwidth]{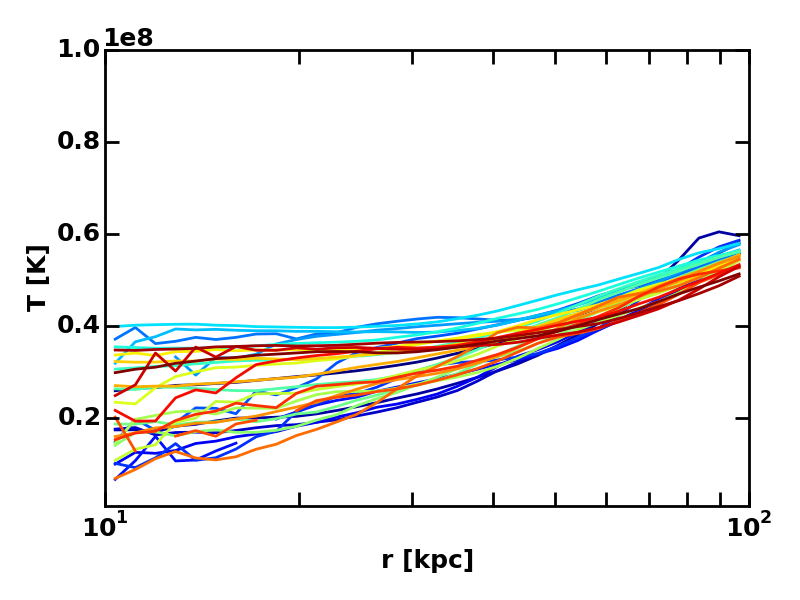}     
         \includegraphics[width=0.24\textwidth]{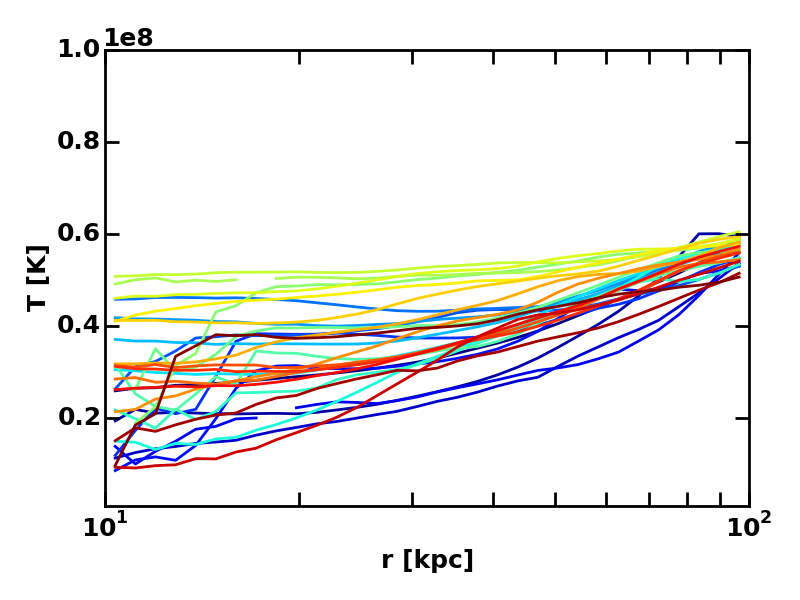}             
                             
      \includegraphics[width=0.24\textwidth]{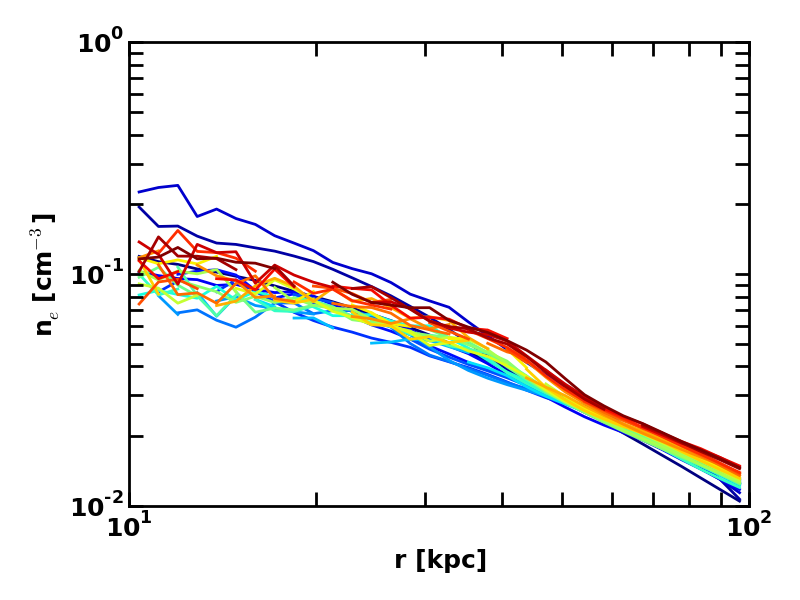}
        \includegraphics[width=0.24\textwidth]{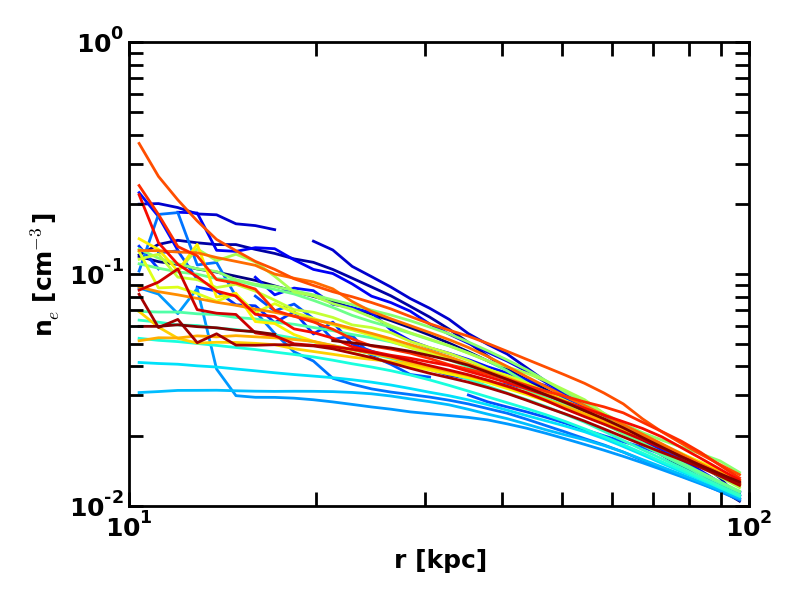} 
         \includegraphics[width=0.24\textwidth]{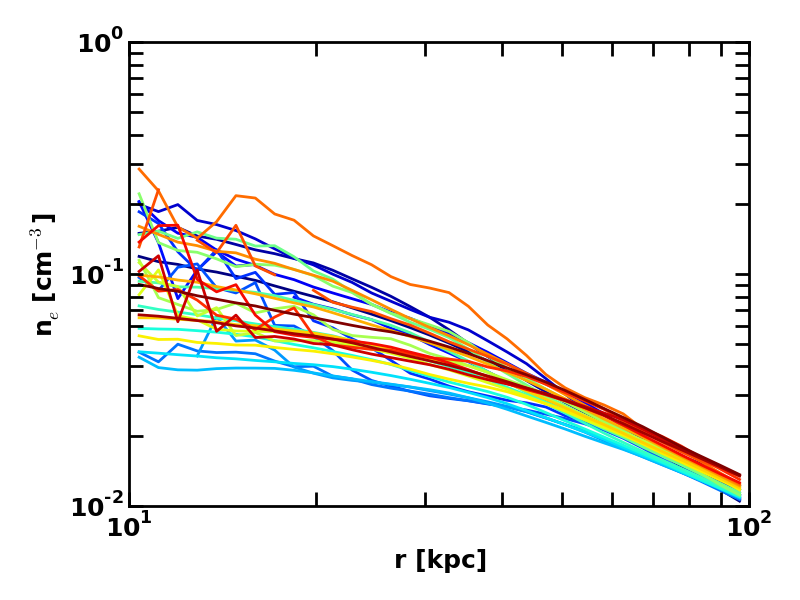}    
        \includegraphics[width=0.24\textwidth]{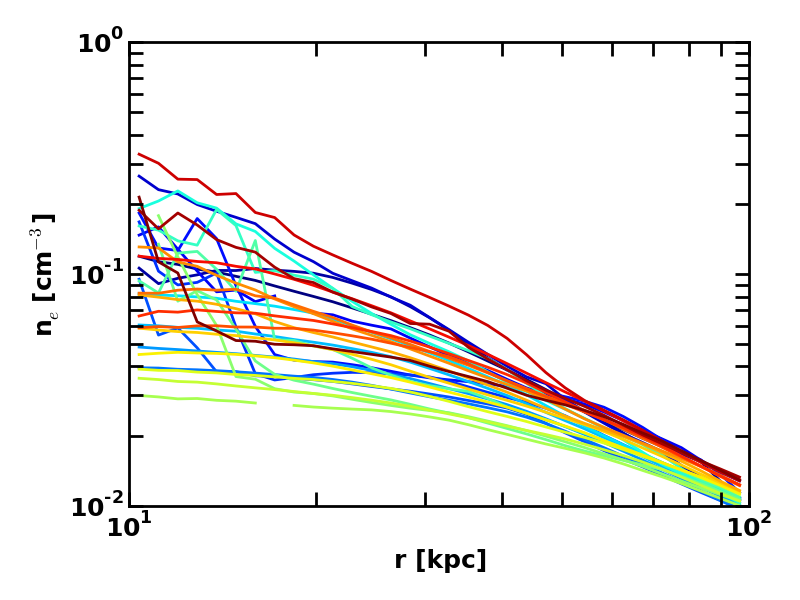}     
                
      \includegraphics[width=0.24\textwidth]{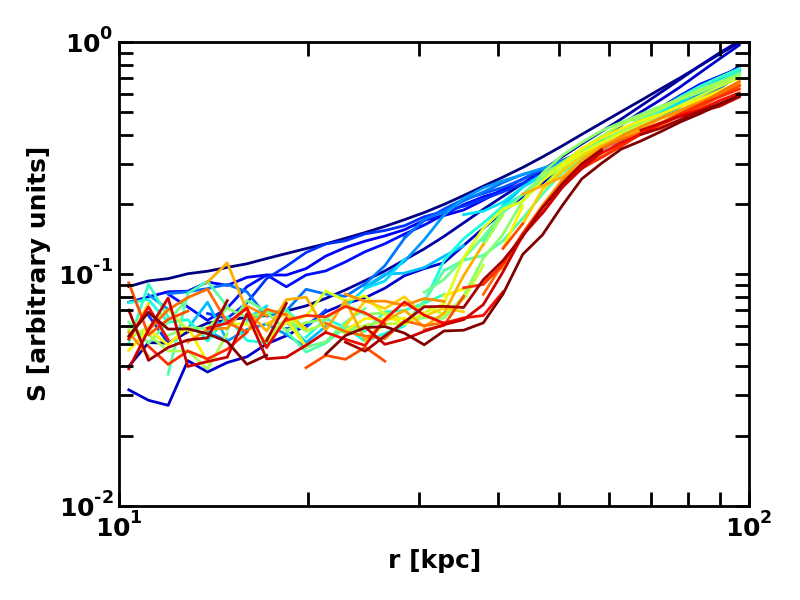}
        \includegraphics[width=0.24\textwidth]{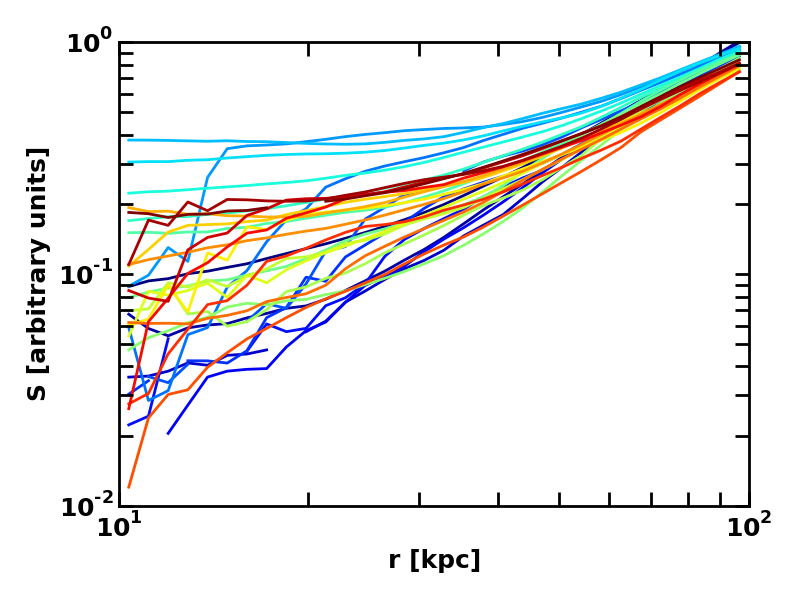} 
        \includegraphics[width=0.24\textwidth]{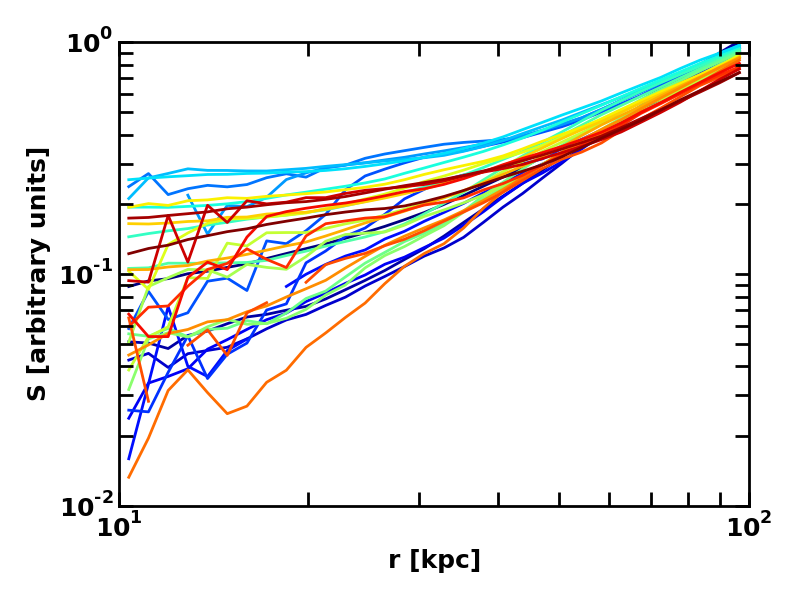}              
        \includegraphics[width=0.24\textwidth]{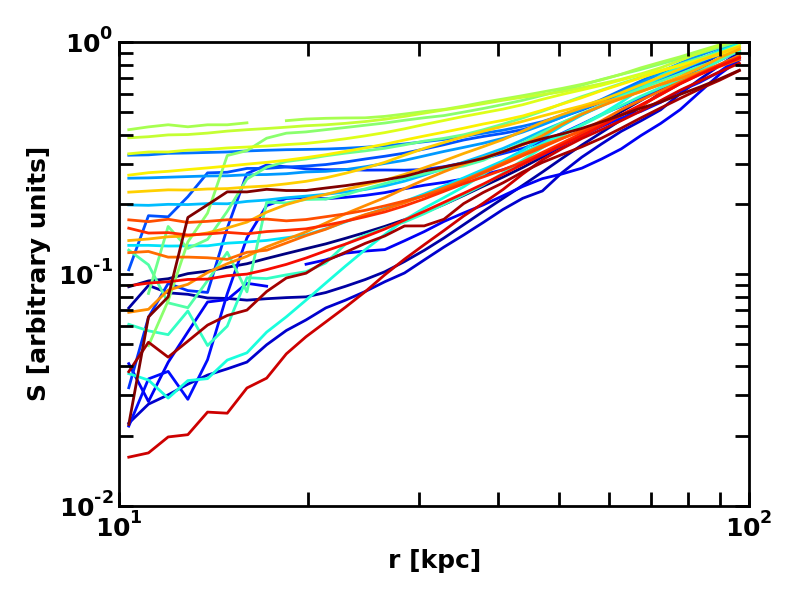}  
        
        \begin{picture}(0,0)
	\put(-230,450){\includegraphics[width=0.1\textwidth]{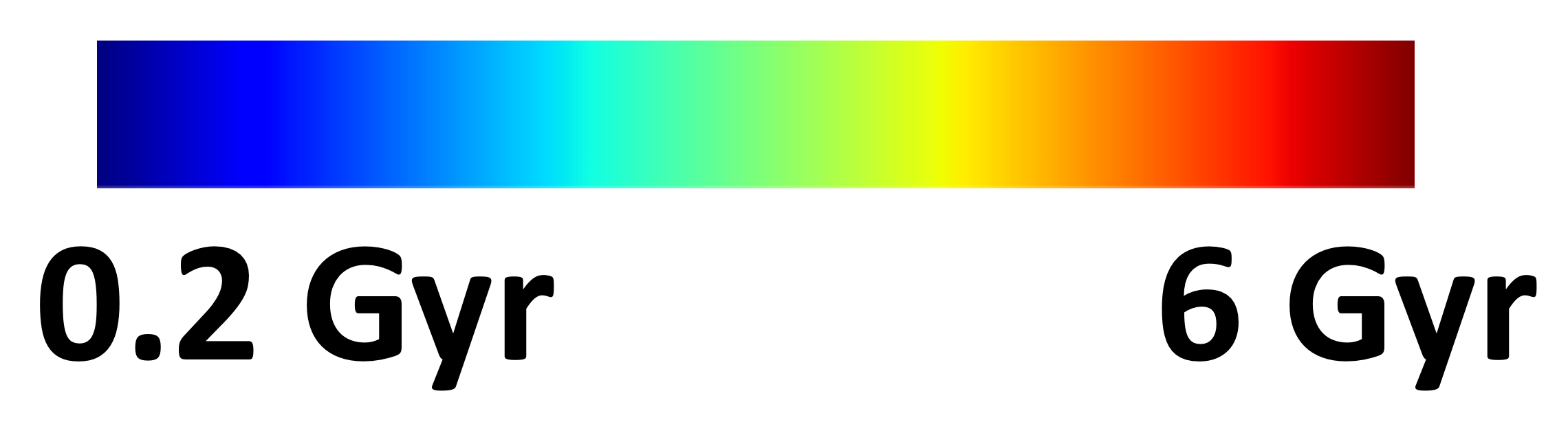}}
	\end{picture}
                         
       \caption[]{Profiles of temperature, entropy normalized to the initial entropy distribution, emission-weighted temperature, emission-weighted density, emission-weighted entropy (from top to bottom, respectively; ordering of columns is the same as in Fig. 1).}
     \label{fig:dens}
  \end{center}
\end{figure*}
\indent
We point out that the increase of the ICM entropy in cool cores may be dominated by CR heating rather than by, for example, turbulent dissipation. After the ICM has come into contact with CRs and experienced localized heating, it can expand locally. Such generated gas motions could eventually decay via turbulent dissipation. However, the primary heating mechanism in this case would be the CR heating rather then ``secondary'' turbulent dissipation. However, we also note that the framework we are using does not allow for the dissipation of sound waves by conductive and viscous processes. While these processes are likely to play an important role too (see, e.g., \citet{Ruszkowski2004a, Ruszkowski2004b, Fabian2017}), including these processes is beyond the scope of this paper. \\
\indent
Typical patterns in the evolution of the heating-to-cooling ratios shown in Figure 5 are dramatically different when CR streaming is included, i.e., in all other panels except for the bottom left panel. It is evident that including streaming increases temporal variability in the CR heating profiles. This variable behavior also mirrors what is seen in Figure 2 showing the evolution of the CR pressure support. In particular, the top left panel in Figure 5, that includes CR streaming and associated with it streaming heating, shows that the source is highly intermittent and that CR heating no longer systematically increases over time. Importantly, each significant AGN outburst results in CR heating rates being comparable to radiative cooling. Similar conclusion can be drawn from the top middle panel that corresponds to the cases that also includes hadronic and Coulomb heating. It also applies to its analog shown in the top right panel that corresponds to super-Alfv{\'e}nic streaming though the heating rates are somewhat reduced due to (i) accelerated transport of CRs away from the center of the cool core and (ii) the fact that the heating rate depends on the gradient of CR distribution that is somewhat flatter in this case due to smoother CR distribution. \\
\indent
We can also compare the contributions of CR streaming heating and the combined Coulomb and hadronic losses to the total heating budget by comparing top and bottom panels in the middle and right columns. Top panels show the contribution from the CR streaming case while the bottom ones that due to the sum of Coulomb and hadronic heating. Interestingly, it is the CR streaming heating that dominates in all cases. \\
\indent
In Figure 6 we show profiles of temperature, entropy normalized to the initial entropy distribution, emission-weighted temperature, emission-weighted density, and emission-weighted entropy (from top to bottom, respectively; ordering of columns is the same as in Fig. 1; weighting is computed using X-ray band extending from 0.5 to 10 keV). Color-coded lines correspond to different times. There is significant qualitative difference between the evolution of the temperature profiles in the non-streaming case (upper left panel) and all other cases. In the non-streaming case, the temperature systematically decreases over time due to the development of global thermal instability which origin, as mentioned above, can be traced back to inefficient mixing of CRs with the thermal ICM and thus inefficient heating of the bulk of the ICM. In all other cases but this one, the cluster atmosphere exhibits temperature variations but profiles vary around an average profile that does not exhibit very low temperatures. Similar trends are seen in the second row that shows profiles of the entropy profile normalized to the initial entropy distribution. Only in the non-streaming case does the gas entropy systematically decrease down to very low values. This demonstrates that the case without CR transport is unsuccessful. Very low gas temperatures and entropies would lead to significant line emission and star formation in excess of what is observed in cool cores. The third row shows X-ray emission-weighted temperature profiles. Unlike the temperature distributions shown in the first row, the emission-weighted ones do not show occasional very large departures from the mean profile, and in particular they do not exhibit centrally inverted temperature slopes, which is consistent with observations. Similarly, the emission-weighted gas density distributions shown in the fourth row are well-behaved. As a side comment, note that the simulations by construction start from a state that is out of thermodynamical equilibrium. This means that we do expect larger temperature variations compared to what one could have predicted starting from hydrostatic and thermal equilibrium in the initial state. Finally, the last row shows emission-weighted entropy profiles and demonstrates that the AGN feedback is gentle enough to preserve the positive entropy gradient in agreement with observations.

\section{Summary and conclusions}
We presented simulations of AGN feedback in cluster cool cores including the effects of CRs. Specifically, our simulations include CR injection by AGN jets, 
CR streaming along the magnetic field lines, radiative cooling, CR heating of the ICM via CR streaming instability, Coulomb interactions and hadronic processes. Our conclusions can be summarized as follows.

\begin{itemize} 

\item We presented a numerical proof of concept that CRs supplied to the ICM via an AGN jet can efficiently heat the ICM in a self-regulating fashion. 
This mode of heating does not demonstrably violate observational constraints as only a low level of CR pressure support is needed to offset radiative cooling during the feedback cycle.

\item The emission-weighted temperature and entropy profiles predicted by this model are broadly consistent with the data.

\item CR streaming is an essential ingredient of the model. When CR streaming is neglected, the CRs inside the AGN-inflated bubbles do not efficiently interact with the ambient thermal ICM, which leads to inefficient coupling of the AGN energy to the ICM, global cooling catastrophe, and excessive accumulation of CRs in the center of the cool core. On the other hand, when streaming is included, CRs mix efficiently with the thermal ICM and transfer their energy to the gas via CR streaming heating and  Coulomb and hadronic interactions. 

\item In the simulations that include CR streaming, the AGN jet and the X-ray luminosity of the cool core are intermittent. When CR transport is neglected, feedback loop is broken, AGN power is relatively weakly variable and is not used efficiently to offset cooling. 

\item When CR streaming heating and Coulomb and hadronic heating processes are all included, it is the CR streaming heating that dominates over other CR heating mechanisms.

\end{itemize}

\acknowledgments{M.R. thanks Department of Astronomy at the University of Maryland for hospitality during his sabbatical stay.  M.R. is grateful for the hospitality of the Harvard-Smithsonian Center for Astrophysics and the Astronomy Department at the University of Wisconsin--Madison, which was made possible in part by a generous gift from Julie and Jeff Diermeier. We thank Ellen Zweibel for very useful discussions, and specifically for highlighting the role of Landau damping. MR thanks Brian McNamara, Aneta Siemiginowska, Ralph Kraft, Christine Jones, Bill Forman,  Reinout van Weeren, and Brian Morsony for very helpful conversations. H.Y.K.Y. acknowledges support by NASA through Einstein Postdoctoral Fellowship grant number PF4-150129 awarded by the Chandra X-ray Center, which is operated by the Smithsonian Astrophysical Observatory for NASA. The software used in this work was in part developed by the DOE NNSA-ASC OASCR Flash Center at the University of Chicago. M.R. acknowledges NASA grant NASA ATP 12-ATP12-0017. C.S.R. thanks for the support from the US National Science Foundation under grant AST 1333514.
Simulations were performed on the Pleiades machine at NASA Ames. Data analysis presented in this paper was performed with the publicly available $yt$ visualization software \citep{turk2011}. We are grateful to the $yt$ development team and the $yt$ community for their support.\\}


\bibliography{crclusters}

\end{document}